\documentclass[fleqn,10pt]{wlscirep}
\usepackage[utf8]{inputenc}
\usepackage[T1]{fontenc}
\usepackage{newtxtext,newtxmath}
\usepackage{graphicx}
\usepackage{dcolumn}
\usepackage{bm}
\usepackage{blindtext}
\usepackage{xcolor}

\title{Disorder as a route to topological spin textures in amorphous magnets}
\author[1,2*]{Temuujin Bayaraa}
\author[1]{Subhashree Satapathy}
\author[1,5]{Peter Fischer}
\author[1,3,4]{Frances Hellman}
\author[1,2]{Sinéad M. Griffin}
\affil[1]{Materials Sciences Division, Lawrence Berkeley National Laboratory, Berkeley, California, 94720, USA}
\affil[2]{Molecular Foundry Division, Lawrence Berkeley National Laboratory, Berkeley, California, 94720, USA}
\affil[3]{Department of Physics, University of California, Berkeley, Berkeley, California}
\affil[4]{Department of Materials Science and Engineering, University of California, Berkeley, Berkeley, California}
\affil[5]{Department of Physics, University of California, Santa Cruz, Santa Cruz, California}
\affil[*]{tbayaraa@lbl.gov}

\date{\today}

\begin{abstract}
Topological phases in solids are typically associated with crystalline symmetry and periodic order. Recent work has shown, however, that reciprocal-space topology can persist even in amorphous materials. Whether real-space topology, encoded directly in magnetic textures, can emerge without crystalline symmetry remains unresolved. Here, we show that structural disorder itself can generate the interactions required to stabilize topological spin textures in amorphous FeGe. Using first-principles calculations to construct a statistical magnetic Hamiltonian, we find that local structural asymmetry and broad distributions of exchange couplings produce a frustrated magnetic network with net ferromagnetic character rather than spin-glass freezing, supporting vortex-, antivortex-, and skyrmion-like textures that self-organize into emergent spin superstructures. Monte Carlo simulations reproduce the experimental Curie temperature and show that these textures persist even when Dzyaloshinskii--Moriya and anisotropy terms are suppressed. In contrast, homogenizing the Hamiltonian strongly suppresses the topological response, identifying bond-to-bond heterogeneity as the essential ingredient. Our results show that real-space topology in a magnet need not be inherited from crystalline chirality, but can emerge from disorder itself.
\end{abstract}

\begin{document}

\maketitle

\section*{Introduction}

Topology in solids is usually formulated in crystalline terms, with symmetry and periodicity organizing the allowed electronic and magnetic states. Recent work on amorphous and quasicrystalline matter has shown, however, that topological phases need not rely on translational order and can even be induced by structural disorder in electronic systems~\cite{corbae2023,Agarwala_et_al:2017,Cain_et_al:2020,corbae2021}. Whether disorder can likewise generate topology carried directly by a spin texture remains unresolved.

Magnetic skyrmions are a real-space form of this where the topology is encoded in the winding of the spin field rather than in a band invariant. In crystalline chiral magnets such as B20 FeGe, they are typically stabilized by Dzyaloshinskii--Moriya interactions generated by spin--orbit coupling and broken inversion symmetry~\cite{muhlbauer2009,Yu2010,Nagaosa2013}. Yet recent experiments have reported skyrmion-like and bubble-like textures in amorphous and nanocrystalline Fe--Ge alloys, in some cases persisting to room temperature~\cite{streubel2021,singh2023}. These observations raise a central question: what stabilizes real-space topology when crystalline chirality is absent?

Here we show that, in amorphous FeGe, structural disorder can itself generate the interactions needed for topological spin textures. We find that exchange depends not only on interatomic separation but also on local moment amplitudes, producing broad distributions of competing ferromagnetic and antiferromagnetic couplings. The resulting frustrated exchange network reproduces the experimental Curie temperature and stabilizes skyrmion-like textures even when Dzyaloshinskii--Moriya and anisotropy terms are excluded. Disorder therefore does more than tolerate topology: it organizes it.

Using amorphous FeGe as an exemplar, we combine \textit{ab initio} modelling, bond-resolved extraction of magnetic interactions, and Monte Carlo simulations to trace the route from structural heterogeneity to frustrated exchange and, ultimately, to topological spin textures in a disordered magnet with net ferromagnetic character rather than spin-glass freezing. More broadly, our results show that real-space topology in a magnet need not be inherited from crystalline chirality, but can emerge from disorder itself.

\section*{Results}

\subsection*{Structural disorder broadens local moments}

Crystalline FeGe adopts the non-centrosymmetric B20 structure, in which Fe sites are symmetry equivalent with uniform local moments. Amorphous FeGe, by contrast, lacks a unique Fe environment: the loss of translational order broadens both local coordination and local moment amplitude, providing the microscopic origin of its structural and magnetic heterogeneity. \textit{Ab initio} melt--quench structures reproduce the experimental density and capture the broad distribution of Fe--Fe and Fe--Ge environments expected for a chemically disordered metallic network (Supplementary Information).

The local magnetic moments reflect this structural diversity. Across the amorphous network, Fe moments span a broad range, from $|m_i|\!\approx\!0$ to $\sim$2.5~$\mu_\mathrm{B}$ (Figs.~\ref{fig1}(d) and S4). As shown in Fig.~\ref{fig2}, the moment amplitude increases approximately linearly with the number of Fe neighbours and decreases with increasing Ge coordination. This trend is consistent with hyperfine measurements in amorphous Fe$_x$Si$_{1-x}$ alloys~\cite{marchal1976,karel2014} and with the picture that moment formation in metallic glasses is controlled by the local density of magnetic neighbours~\cite{coey1978}. 

The sign of the local moments also correlates nearly linearly with the mean magnetic field generated by neighbouring spins (Fig.~\ref{fig2}(c)), indicating a strong local mean-field alignment tendency.

Amorphous FeGe therefore hosts a structurally encoded distribution of local moment amplitudes, providing the microscopic basis for the heterogeneous magnetic interactions discussed below.

\subsection*{Exchange heterogeneity frustrates the amorphous magnet}

The structural heterogeneity in the amorphous network supports a dense manifold of low-lying magnetic states, so no single ferromagnetic or antiferromagnetic initialization provides a representative description of its magnetism. We therefore relaxed 71 randomized noncollinear spin initializations for the same underlying amorphous structure (Methods). The resulting atomic rearrangements were minimal, with supercell dimensions changing by less than 2\% and the volume by less than 0.8\%, indicating that the sampled states remain within a single structural basin.

The resulting ensemble spans average Fe moments from 0.08 to 1.56~$\mu_\mathrm{B}$ per Fe atom, bracketing the experimental value of 0.75~$\mu_\mathrm{B}$~\cite{Bouma2020}. Lower-energy states generally carry larger average magnetization, and several energetically favourable configurations fall within the experimental range. Experimentally relevant moments, therefore, emerge naturally once the noncollinear magnetic landscape is sampled rather than imposed. A simple Boltzmann weighting of the sampled DFT states illustrates how thermal occupation of nearby magnetic minima can renormalize the apparent moment in an amorphous magnet (Fig.~\ref{fig1}(c)). We use this only as a qualitative illustration; quantitative finite-temperature behaviour is obtained from Monte Carlo simulations of the \textit{ab initio}-derived spin Hamiltonian below.

We next extract a bond-resolved spin Hamiltonian for the lowest-energy amorphous configuration identified in Fig.~\ref{fig1}(b), including exchange couplings ($J_{ij}$), Dzyaloshinskii--Moriya interaction (DMI) vectors ($\mathbf{D}_{ij}$), and single-ion anisotropy (SIA) tensors ($\mathbf{A}_i$). The interactions are short-ranged: both exchange and DMI decay rapidly beyond 3.5~\AA{} (Fig.~\ref{fig3}). Throughout, negative $J_{ij}$ denotes ferromagnetic coupling.

The exchange interactions span a strikingly broad range in both magnitude and sign, from strongly ferromagnetic to antiferromagnetic values (Fig.~\ref{fig3}(a)). Roughly threequarters of the bonds are ferromagnetic and one quarter antiferromagnetic, consistent with a net ferromagnetic tendency built from strongly heterogeneous local couplings.

Exchange in amorphous FeGe is not a single-valued function of interatomic distance. Earlier first-principles studies of amorphous alloys reported substantial scatter around a single $J(r)$ curve~\cite{terasawa2020,Lindmaa2013}, but did not explicitly treat moment amplitude as a separate local descriptor of exchange. Bonds of similar length can carry very different couplings depending on the amplitudes of the connected local moments, showing that $J_{ij}$ depends jointly on geometry and moment magnitude. A compact surrogate model based on $r_{ij}$ and $m_i m_j$ captures the dominant trends (Supplementary Fig.~S11), demonstrating that moment heterogeneity is not merely scatter around a distance-only trend, but an essential descriptor of exchange variability.

Short Fe--Fe bonds connecting strongly magnetic sites tend to support robust ferromagnetic exchange, whereas longer or Ge-mediated pathways, often involving reduced local moments, yield weaker or antiferromagnetic couplings. DMI and SIA are likewise distributed across the amorphous network, reflecting local inversion breaking and coordination-dependent anisotropy, but their characteristic energy scales remain well below those of exchange. The defining feature of the amorphous Hamiltonian is therefore heterogeneous short-range exchange rather than large spin--orbit interactions.

The coexistence of strong ferromagnetic and antiferromagnetic bonds makes the network intrinsically frustrated: no single spin configuration can satisfy all local interactions simultaneously. This frustration generates strongly noncollinear spin correlations and locally twisting spin fields, providing the microscopic seed of the topological textures described below. Amorphous FeGe thus realizes a route to topological spin textures in which topology emerges from structural disorder and exchange heterogeneity rather than from symmetry-constrained interactions.

\subsection*{The heterogeneous Hamiltonian stabilizes topological textures}
We next examine the collective magnetic states that emerge from this heterogeneous exchange network. Monte Carlo simulations of the bond-resolved amorphous spin Hamiltonian yield a Curie temperature $T_\mathrm{C}\!\approx\!325$~K (Fig.\ref{fig4}(a)), close to the experimental value of $\sim$300~K~\cite{singh2023}. The magnetization decreases gradually with temperature, reflecting the broad spatial distribution of local interaction strengths characteristic of the amorphous network (Fig.\ref{fig4}(c)).

Near room temperature, the simulated configurations are strongly noncollinear and carry finite topological charge, including vortex-, antivortex-, and skyrmion-like textures (Fig.~\ref{fig4}). The characteristic cores span volumes of roughly 25--200~\AA$^3$ with typical separations of 6--15~\AA. Larger simulated slices reveal multiple ferromagnetic regions converging into vortex centres, consistent with the coexistence of several local topological objects within the amorphous structure. The extracted Hamiltonian therefore supports topologically nontrivial real-space textures despite the absence of crystalline symmetry.

To assess the relative importance of different magnetic interactions, we evaluated their contributions to the total magnetic energy of the simulated $4\times4\times4$ amorphous supercell. Exchange overwhelmingly dominates the magnetic energetics, exceeding the DMI and SIA contributions by several orders of magnitude ($-5.39$, $-0.006$, and $-0.001$~eV, respectively). Spin--orbit-derived terms therefore refine the magnetic textures rather than determine their existence.

This interpretation is confirmed by simulations with DMI and SIA set to zero. In that limit, noncollinear textures remain stable, although the density of topological charge is reduced. This shows that spin--orbit interactions are not required for texture formation in the amorphous phase, consistent with recent experimental observations of skyrmion-like textures in amorphous Fe$_{78}$Si$_9$B$_{13}$ and related systems~\cite{wu2023_amorphous}.

The crucial control is obtained by homogenizing the amorphous Hamiltonian while leaving the atomic geometry unchanged. When the bond-resolved exchange and DMI terms are replaced by their mean or median values, the Curie temperature shifts to $\approx$390~K and $\approx$250~K, respectively, and nonzero topological charge becomes rare except in a narrow window near $T_\mathrm{C}$. The disordered lattice alone is therefore not sufficient: the observed textures require the full bond-to-bond heterogeneity of the amorphous Hamiltonian.

Amorphous FeGe therefore shows that topology arises not despite structural disorder but because of it, through the coupling of structural heterogeneity, exchange frustration, and emergent noncollinear spin correlations.

\section*{Discussion}

Recent experiments have shown that amorphous FeGe and related alloys can host skyrmion-like spin textures despite lacking crystalline order~\cite{streubel2021,singh2023,wu2023_amorphous}. Our results show that these textures arise naturally from the heterogeneous exchange interactions and magnetic frustration inherent to the amorphous structure, rather than primarily from the symmetry-driven spin--orbit mechanisms that stabilize skyrmions in crystalline chiral magnets. This provides a microscopic framework for understanding topological magnetism in structurally disordered systems, in which noncollinear textures emerge from the interplay of moment fluctuations, geometric randomness, and competing short-range exchange interactions.

In crystalline magnets, chiral spin states such as skyrmions and helices are usually stabilized either by the Dzyaloshinskii--Moriya interaction or by frustrated exchange acting on a periodic lattice~\cite{muhlbauer2009,Yu2010,Nagaosa2013,Okubo_et_al:2012,Leonov2015,Kurumaji2019}. Amorphous FeGe follows a different route. The absence of translational symmetry does not eliminate the ingredients for topology; instead, it redistributes them locally. Each Fe site experiences a distinct coordination environment and moment amplitude, producing a broad spectrum of exchange couplings $J_{ij}$ that vary in both sign and strength. This bond-to-bond variability generates intrinsic magnetic frustration, creating locally correlated regions whose mutual incompatibility stabilizes vortex-, antivortex-, and skyrmion-like textures even without a global chiral axis.

The relation $J_{ij}=J(r_{ij},m_i,m_j)$ identified here highlights the central role of moment fluctuations in amorphous magnetism. When structural and magnetic heterogeneity occur on comparable energy scales, a distance-only Heisenberg description becomes insufficient: bonds of similar length need not carry similar exchange because the local moment amplitudes also encode the surrounding chemical environment. Topological textures in the amorphous phase are therefore not a residual consequence of crystalline-like interactions broadened by disorder, but the collective outcome of a genuinely heterogeneous magnetic network. In this sense, topology emerges from the distribution of local exchange pathways, making it a fundamentally real-space mechanism.

This interpretation is further sharpened by the control calculations. First, noncollinear textures persist when DMI and single-ion anisotropy are removed, showing that spin--orbit-derived terms refine rather than create the effect. Second, homogenizing the Hamiltonian strongly suppresses the topological response, demonstrating that topology does not arise from finite exchange alone but from the full distribution of local couplings. The essential ingredient is thus not merely the presence of competing interactions, but their bond-to-bond variability across the amorphous network. This conclusion is consistent with broader theoretical work showing that exchange frustration can substantially enhance skyrmion stability even in systems with a ferromagnetic ground state~\cite{vonMalottki2017}.

More broadly, these results suggest that disorder itself can act as a design principle for topological magnetism. In crystalline materials, the existence of chiral spin textures is tightly constrained by lattice symmetry and by the availability of suitable spin--orbit interactions~\cite{Nagaosa2013}. Amorphous magnets, by contrast, naturally generate heterogeneous exchange networks whose frustration can stabilize noncollinear spin states without requiring a symmetry-imposed handedness. This widens the conceptual and materials landscape for topological magnetism: rather than searching only for crystals with the correct symmetry, one can ask how local coordination statistics, chemical disorder, and moment variability can be engineered to favour real-space topology. In that respect, amorphous FeGe points to a route that is distinct from, but complementary to, dipole-stabilized topological textures reported in other amorphous systems such as Fe/Gd multilayers~\cite{Montoya2017}.

The exchange-dominated character of the amorphous Hamiltonian also suggests a route to topological textures in systems that do not rely primarily on strong relativistic interactions. That possibility is appealing because it points beyond the usual heavy-element design rules of chiral magnetism and toward a broader class of disordered materials in which local exchange frustration provides the dominant organizing principle. If such textures can be stabilized and controlled reproducibly, amorphous magnets may offer a flexible platform for topological spin textures in thin-film geometries, where the absence of lattice-matching constraints is often an advantage rather than a limitation.

Our framework, therefore, provides a natural interpretation of the growing body of experimental reports of chiral spin correlations in amorphous alloys~\cite{streubel2021,singh2023,wu2023_amorphous}. The key ingredient for topology is not crystalline chirality itself, but the coexistence of local asymmetry, frustrated exchange, and moment disorder. Future work combining real-space magnetic imaging, element-sensitive probes, and large-scale spin-dynamics simulations could test how these amorphous textures respond to field, strain, confinement, and current, and could determine the extent to which their topology is robust, mobile, and controllable.

In summary, amorphous FeGe shows that topology can arise not despite structural disorder but because of it. By identifying exchange heterogeneity as the microscopic origin of skyrmion-like textures in the amorphous phase, this work extends topological magnetism beyond ordered crystals and demonstrates how structurally disordered magnets as a platform in which local asymmetry, moment fluctuations, and frustrated exchange can generate robust real-space topology.

\begin{figure}
\includegraphics[width=1\textwidth]{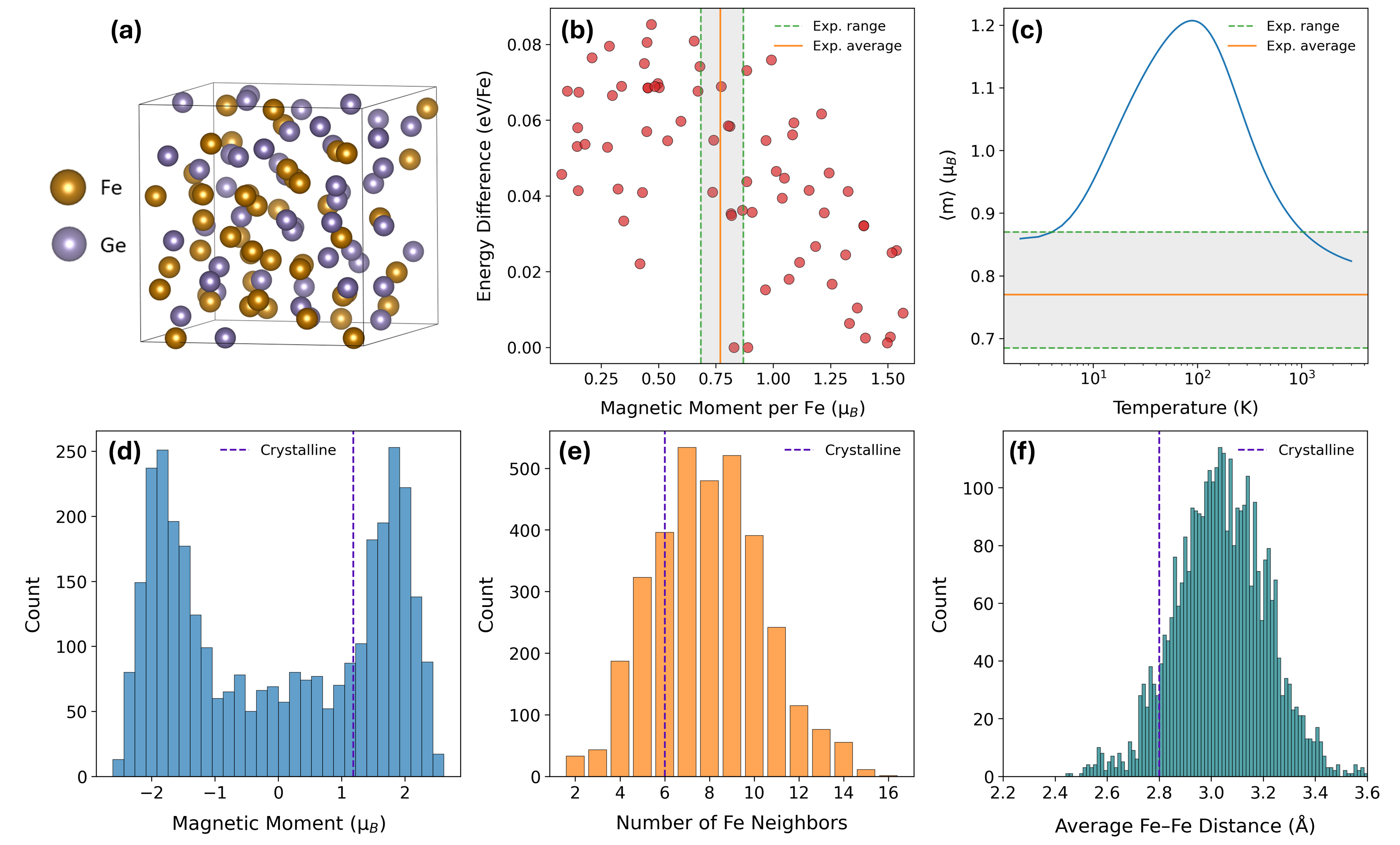}
\centering
\caption{\textbf{Magnetic landscape of amorphous FeGe.}
\textbf{a} Atomic structure of the amorphous FeGe supercell obtained from \textit{ab initio} melt–quench simulations. Fe atoms are shown in orange and Ge atoms in blue. \textbf{b} Calculated total energies of 71 relaxed noncollinear magnetic configurations obtained from different spin initializations of the same amorphous atomic structure. \textbf{c} Boltzmann-weighted estimate of the thermally averaged magnetic moment obtained from the ensemble of DFT states, illustrating how thermal occupation of nearby magnetic minima can renormalize the observed moment in an amorphous magnet. The shaded region, green and orange lines are derived from experimental values reported in Ref.~\cite{Bouma2020}. \textbf{d-f} Distribution of the local Fe magnetic moments, the number of Fe neighbours, and the average Fe--Fe distance. Our analysis was performed within a sphere of 3.5\AA~radius. Vertical dashed lines represent references from the B20 crystal FeGe structure. Negative magnetic moments are with respect to the global magnetization axis. Together, these results reveal a dense manifold of low-lying magnetic states in amorphous FeGe arising from structural disorder.}
\label{fig1}
\end{figure}

\begin{figure}
\includegraphics[width=0.8\textwidth]{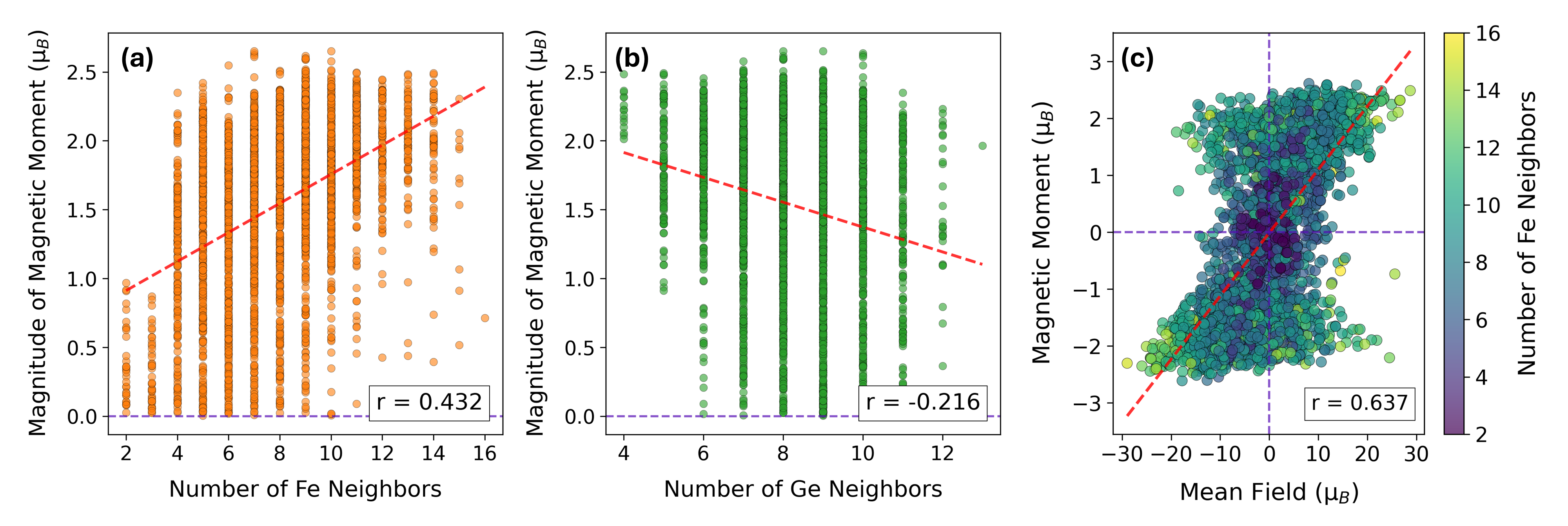}
\centering
\caption{\textbf{Local coordination controls the magnitude of Fe magnetic moments in amorphous FeGe.} \textbf{a} Magnitude of the local Fe moment, $|m_i|$, as a function of the number of Fe neighbours within 3.5\,\AA. Fe-rich coordination environments stabilize larger magnetic moments. \textbf{b} $|m_i|$ versus the number of Ge neighbours within 3.5\,\AA. Increasing Ge coordination suppresses the Fe moment, consistent with enhanced Fe–Ge hybridization. \textbf{c} $|m_i|$ versus the magnitude of the local mean field, $h_i = \sum_i m_{i}$ where neighbours are considered within cutoff 3.5\,\AA. Stronger local mean fields correlate with larger magnetic moments. Solid lines are linear fits to the data, and inset annotations report the fit slope values. Together, these trends demonstrate that structural disorder produces a broad distribution of local moment amplitudes, providing the microscopic origin of the heterogeneous magnetic interactions discussed in the main text.}
\label{fig2}
\end{figure}

\begin{figure}
\includegraphics[width=1\textwidth]{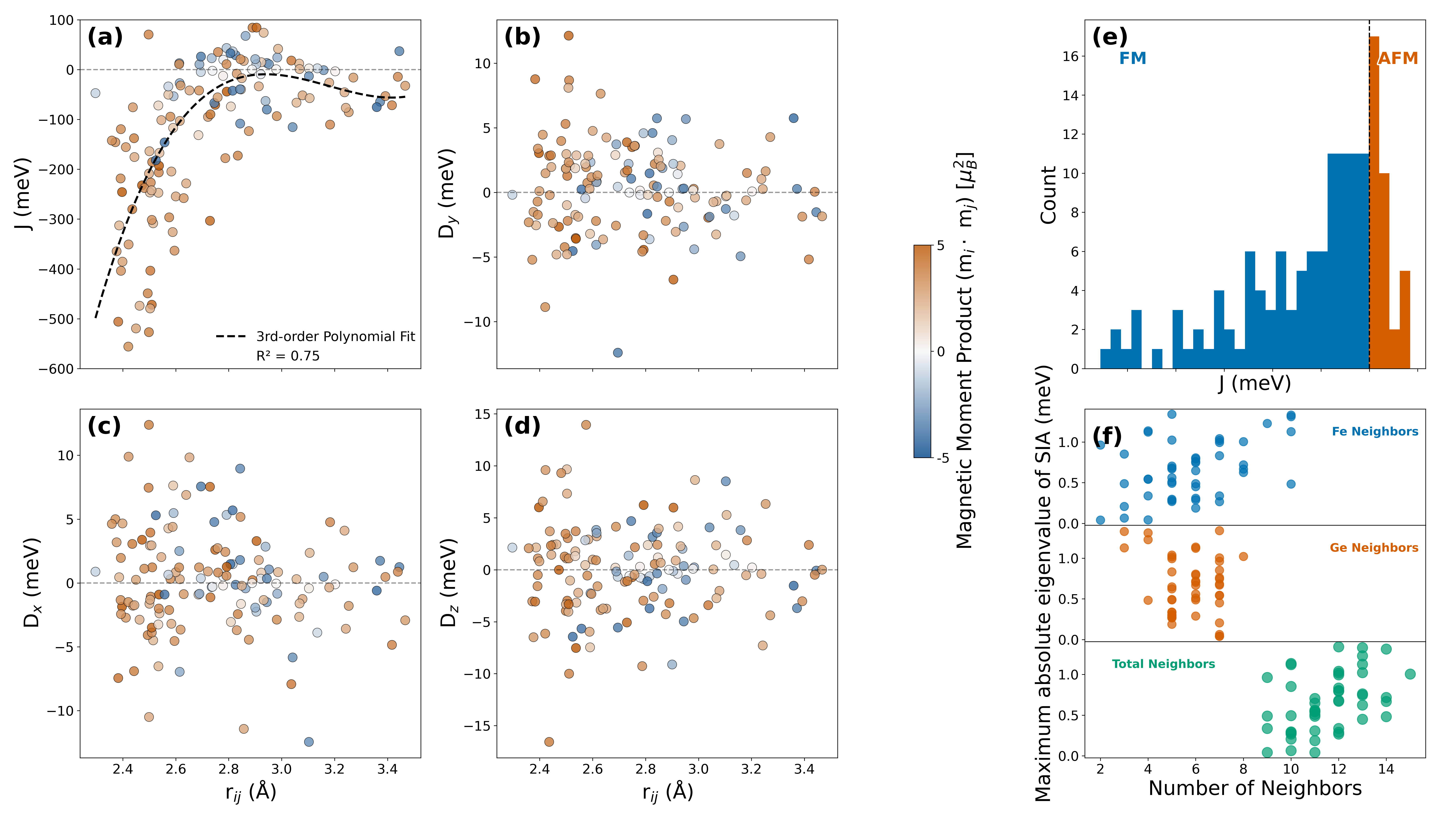}
\caption{\textbf{Heterogeneous magnetic interactions in amorphous FeGe.}
\textbf{a–d} Magnetic exchange couplings $J_{ij}$ and the three components of the Dzyaloshinskii–Moriya interaction (DMI) vectors as a function of Fe–Fe pair distance. Colour maps indicate the product of the local magnetic moments $m_i m_j$ associated with each pair. Interactions decay rapidly beyond $\sim$3.5\,\AA. Negative $J_{ij}$ denotes ferromagnetic exchange. We fitted $J$ using multivariate polynomial regression $J_{fit} = \sum_{i=0}^{N} \sum_{j=0}^{N-i} c_{ij} r^i |\mu|^j$, where $r$ is the pair distance and $\mu$ is the product of the magnetic moments of the pairs, and found that the model fits well to the $J$ values with $R^2$ reaching 0.75 with 3rd-order polynomial degree. \textbf{e} Distribution of exchange interactions across the amorphous network, showing a broad spread of both ferromagnetic and antiferromagnetic couplings. \textbf{f} Magnitude of the single-ion anisotropy (SIA) on Fe sites as a function of local coordination, shown versus the number of Fe, Ge, and total nearest neighbours. Together, these results demonstrate that amorphous FeGe hosts a highly heterogeneous magnetic Hamiltonian in which exchange interactions vary strongly with both local geometry and moment amplitude, while spin–orbit-derived interactions remain much smaller in magnitude.}
\label{fig3}
\end{figure}

\begin{figure*}
\includegraphics[width=1\textwidth]{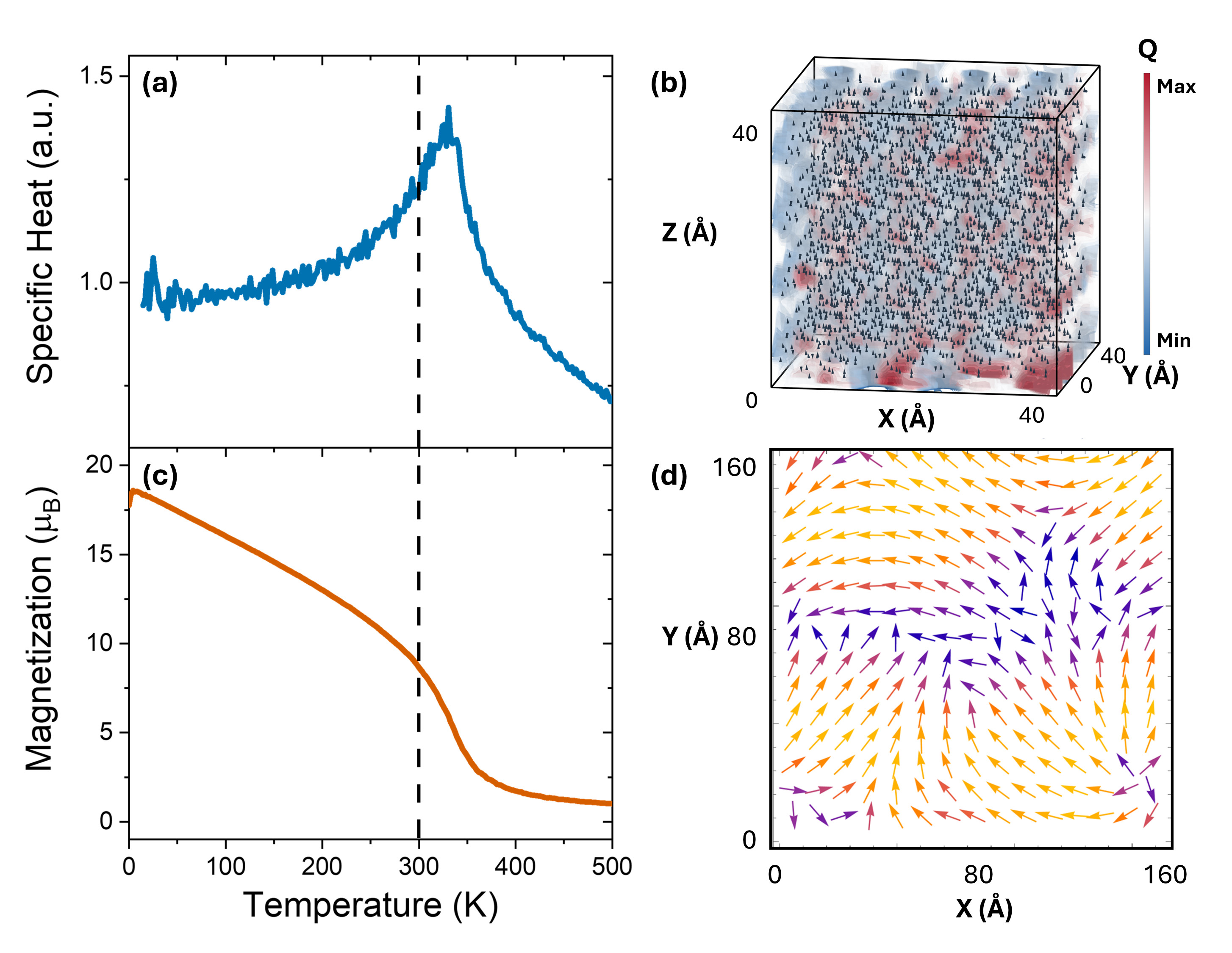}
\caption{\textbf{Magnetic properties and emergent spin textures in amorphous FeGe from Monte Carlo simulations.} \textbf{a} Specific heat (arbitrary units) as a function of temperature, showing a peak near the magnetic transition. \textbf{b} Distribution of topological charge $Q$ and representative spin texture of the amorphous $4\times4\times4$ supercell obtained from Monte Carlo simulations at room temperature. Spins are shown as grey cones at each magnetic site.  \textbf{c} Magnetization as a function of temperature, yielding a Curie temperature consistent with experimental measurements.  \textbf{d} Two-dimensional vector field of spins in an amorphous $16\times16\times1$ supercell at room temperature, where spin directions are coloured according to the in-plane angle. Vectors represent the average spin of 48 Fe atoms. The configuration reveals vortex- and skyrmion-like spin textures emerging from the heterogeneous magnetic interactions. Vertical dashed lines in panels \textbf{a} and \textbf{c} indicate the Curie temperature obtained from experimental Curie–Weiss fits reported in Ref.~\cite{singh2023}. Together, these simulations show that heterogeneous short-range exchange interactions can stabilize topological spin textures even in the absence of crystalline symmetry.}
\label{fig4}
\end{figure*}

\section*{Methods}

\subsection*{Density functional theory calculations}

All electronic structure calculations were performed within density functional theory using the Vienna \textit{Ab Initio} Simulation Package (VASP) \cite{Kresse1996,Kresse1999}. The projector augmented-wave (PAW) method was used to describe the ionic cores \cite{Blochl1994}. Exchange–correlation effects were treated using the generalized gradient approximation in the Perdew–Burke–Ernzerhof form \cite{Perdew1992}.

Electronic wavefunctions were expanded in a plane-wave basis with a kinetic energy cutoff of 500~eV. Brillouin-zone sampling used Monkhorst–Pack grids for crystalline calculations, while amorphous supercells were sampled at the $\Gamma$ point due to their large real-space dimensions.

Spin–orbit coupling was included in all DFT calculations.

Local magnetic moments were obtained directly from the self-consistent spin density projected onto atomic spheres. Because amorphous structures exhibit substantial variation in local atomic environments, no fixed Wigner–Seitz radius was imposed.

\subsection*{Generation of amorphous FeGe structures}

Amorphous FeGe structures were generated using an \textit{ab initio} melt–quench molecular dynamics protocol. A 100-atom supercell with stoichiometry Fe$_{50}$Ge$_{50}$ was initialized in a random configuration and heated above the melting temperature to produce a liquid phase. The liquid was equilibrated and subsequently quenched to room temperature.

Following the quench, all atomic positions were fully relaxed until residual forces were below $10^{-2}$ eV/\AA. The resulting atomic density agrees with experimentally reported densities of sputtered amorphous FeGe films within approximately 3\%.

The final structures exhibit the short- and medium-range order typical of metallic glasses. Fe–Fe and Fe–Ge bond lengths span approximately $2.3$–$3.1$ \AA, and coordination numbers vary between roughly 6 and 10 depending on the local atomic environment (Figs.~\ref{fig1} and S4).

\subsection*{Construction of the spin Hamiltonian}

Magnetic interactions were obtained by mapping the DFT results onto an atomistic spin Hamiltonian

\begin{equation}
H =
\sum_{<i,j>}^{NNs} J_{ij} \boldsymbol{S_i} \cdot \boldsymbol{S_j}
+ \sum_{<i,j>}^{NNs} \boldsymbol{D} \cdot \boldsymbol{S_i} \times \boldsymbol{S_j}
+ \sum_i \boldsymbol{S_i} \cdot \boldsymbol{A_{ii}} \cdot \boldsymbol{S_i} .
\end{equation}

Here $J_{ij}$ denotes the magnetic exchange coupling between spins $i$ and $j$, $\mathbf{D}_{ij}$ is the Dzyaloshinskii–Moriya interaction (DMI) vector, and $A_i$ is the single-ion anisotropy tensor.

Interactions were extracted for all Fe–Fe pairs within a cutoff radius of $3.5$ \AA, yielding 139 magnetic bonds in the amorphous structure. All magnetic parameters are obtained using the 4-state energy mapping method \cite{Xiang2013,Xu2019,Bayaraa2021a} and the Green's function method \cite{tb2j} for the amorphous and crystalline structures, respectively.

The extracted exchange interactions span a wide range of magnitudes and signs, reflecting the diversity of local atomic environments in the amorphous network.

\subsection*{Monte Carlo simulations}

Finite-temperature magnetic behaviour was investigated using classical Monte Carlo simulations \cite{Miyatake1986,PASP} of the DFT-derived spin Hamiltonian. Simulations employed the Metropolis algorithm on supercells constructed by periodically replicating the amorphous structure.

Typical calculations used $4 \times 4 \times 4$ supercells containing several thousand spins.
For each temperature, the system was equilibrated for $10^5$ Monte Carlo sweeps followed by an additional $10^5$ sweeps used for statistical averaging.

The Curie temperature was determined from the temperature dependence of the magnetization.

Additional simulations were performed in which DMI and single-ion anisotropy were set to zero to assess their influence on the magnetic textures.

\subsection*{Topological charge analysis}

Topological charge, $Q$, is calculated using the definition of Berg and Lüstcher \cite{Berg1981} for a discrete lattice of spins as $Q = 1/4\pi \sum_l A_l$, where $l$ runs over all elementary triangles made of neighbouring spin sites and $A_l$ is the solid angle formed by the three spin vectors of the triangle. Here, for the neighbouring spin sites, we considered all the neighbours within a 3.5\AA~ radius. However, since we find various Fe neighbour counts ranging from 2 to 16, solid angles were calculated for up to the 8th nearest neighbour (see Fig. S4).

Spatial maps of topological charge were constructed from the Monte Carlo configurations to identify vortex centres and skyrmion cores within the amorphous supercells.

\subsection*{Data availability}

All data corresponding to this work are available at ~\cite{zenodo_doi}.

\vspace{12pt}
\noindent\textbf{Acknowledgements}\\

We thank Sujoy Roy, David Raftrey, Zoey Tumbleson, and Guy Moore for useful discussions. This work is primarily supported by the U.S. Department of Energy, Office of Science, Office of Basic Energy Sciences, Materials Sciences and Engineering Division under Contract No. DE-AC02-05-CH11231 within the Nonequilibrium Magnetic Materials Program (MSMAG). Computational resources were provided by the National Energy Research Scientific Computing Center and the Molecular Foundry, DOE Office of Science User Facilities supported by the Office of Science, U.S. Department of Energy under Contract No. DEAC02-05CH11231. The work performed at the Molecular Foundry was supported by the Office of Science, Office of Basic Energy Sciences, of the U.S. Department of Energy under the same contract.

\bibliography{library}

\begin{thebibliography}{10}
\urlstyle{rm}
\expandafter\ifx\csname url\endcsname\relax
  \def\url#1{\texttt{#1}}\fi
\expandafter\ifx\csname urlprefix\endcsname\relax\def\urlprefix{URL }\fi
\expandafter\ifx\csname doiprefix\endcsname\relax\def\doiprefix{DOI: }\fi
\providecommand{\bibinfo}[2]{#2}
\providecommand{\eprint}[2][]{\url{#2}}

\bibitem{corbae2023}
\bibinfo{author}{Corbae, P.} \emph{et~al.}
\newblock \bibinfo{journal}{\bibinfo{title}{{Observation of spin-momentum locked surface states in amorphous Bi$_2$Se$_3$}}}.
\newblock {\emph{\JournalTitle{Nature Materials}}} \textbf{\bibinfo{volume}{22}}, \bibinfo{pages}{200--206}, \doiprefix\url{10.1038/s41563-022-01458-0} (\bibinfo{year}{2023}).
\newblock \eprint{1910.13412}.

\bibitem{Agarwala_et_al:2017}
\bibinfo{author}{Agarwala, A.} \& \bibinfo{author}{Shenoy, V.~B.}
\newblock \bibinfo{journal}{\bibinfo{title}{Topological insulators in amorphous systems}}.
\newblock {\emph{\JournalTitle{Phys. Rev. Lett.}}} \textbf{\bibinfo{volume}{118}}, \bibinfo{pages}{236402}, \doiprefix\url{10.1103/PhysRevLett.118.236402} (\bibinfo{year}{2017}).

\bibitem{Cain_et_al:2020}
\bibinfo{author}{Cain, J.~D.}, \bibinfo{author}{Azizi, A.}, \bibinfo{author}{Conrad, M.}, \bibinfo{author}{Griffin, S.~M.} \& \bibinfo{author}{Zettl, A.}
\newblock \bibinfo{journal}{\bibinfo{title}{Layer-dependent topological phase in a two-dimensional quasicrystal and approximant}}.
\newblock {\emph{\JournalTitle{Proceedings of the National Academy of Sciences}}} \textbf{\bibinfo{volume}{117}}, \bibinfo{pages}{26135--26140}, \doiprefix\url{10.1073/pnas.2015164117} (\bibinfo{year}{2020}).
\newblock \eprint{https://www.pnas.org/doi/pdf/10.1073/pnas.2015164117}.

\bibitem{corbae2021}
\bibinfo{author}{Corbae, P.}, \bibinfo{author}{Hellman, F.} \& \bibinfo{author}{Griffin, S.~M.}
\newblock \bibinfo{journal}{\bibinfo{title}{Structural disorder-driven topological phase transition in noncentrosymmetric bitei}}.
\newblock {\emph{\JournalTitle{Phys. Rev. B}}} \textbf{\bibinfo{volume}{103}}, \bibinfo{pages}{214203}, \doiprefix\url{10.1103/PhysRevB.103.214203} (\bibinfo{year}{2021}).

\bibitem{muhlbauer2009}
\bibinfo{author}{Muhlbauer, S.} \emph{et~al.}
\newblock \bibinfo{journal}{\bibinfo{title}{{Skyrmion Lattice in a Chiral Magnet}}}.
\newblock {\emph{\JournalTitle{Science}}} \textbf{\bibinfo{volume}{323}}, \bibinfo{pages}{915--919}, \doiprefix\url{10.1126/science.1166767} (\bibinfo{year}{2009}).
\newblock \eprint{0902.1968}.

\bibitem{Yu2010}
\bibinfo{author}{Yu, X.~Z.} \emph{et~al.}
\newblock \bibinfo{journal}{\bibinfo{title}{{Real-space observation of a two-dimensional skyrmion crystal}}}.
\newblock {\emph{\JournalTitle{Nature}}} \textbf{\bibinfo{volume}{465}}, \bibinfo{pages}{901--904}, \doiprefix\url{10.1038/nature09124} (\bibinfo{year}{2010}).

\bibitem{Nagaosa2013}
\bibinfo{author}{Nagaosa, N.} \& \bibinfo{author}{Tokura, Y.}
\newblock \bibinfo{title}{{Topological properties and dynamics of magnetic skyrmions}}, \doiprefix\url{10.1038/nnano.2013.243} (\bibinfo{year}{2013}).

\bibitem{streubel2021}
\bibinfo{author}{Streubel, R.} \emph{et~al.}
\newblock \bibinfo{journal}{\bibinfo{title}{{Chiral Spin Textures in Amorphous Iron–Germanium Thick Films}}}.
\newblock {\emph{\JournalTitle{Advanced Materials}}} \textbf{\bibinfo{volume}{33}}, \bibinfo{pages}{2004830}, \doiprefix\url{10.1002/adma.202004830} (\bibinfo{year}{2021}).

\bibitem{singh2023}
\bibinfo{author}{Singh, A.} \emph{et~al.}
\newblock \bibinfo{journal}{\bibinfo{title}{{Characterizing Temporal Heterogeneity by Quantifying Nanoscale Fluctuations in Amorphous Fe‐Ge Magnetic Films}}}.
\newblock {\emph{\JournalTitle{Advanced Functional Materials}}} \textbf{\bibinfo{volume}{33}}, \bibinfo{pages}{2300224}, \doiprefix\url{10.1002/adfm.202300224} (\bibinfo{year}{2023}).

\bibitem{marchal1976}
\bibinfo{author}{Marchal, G.}, \bibinfo{author}{Mangin, P.}, \bibinfo{author}{Piecuch, M.} \& \bibinfo{author}{Janot, C.}
\newblock \bibinfo{journal}{\bibinfo{title}{{M{"{o}}ssbauer study of magnetic ordering in amorphous Fe-Si alloys}}}.
\newblock {\emph{\JournalTitle{Le Journal de Physique Colloques}}} \textbf{\bibinfo{volume}{37}}, \bibinfo{pages}{C6--763--C6--768}, \doiprefix\url{10.1051/jphyscol:19766160} (\bibinfo{year}{1976}).

\bibitem{karel2014}
\bibinfo{author}{Karel, J.} \emph{et~al.}
\newblock \bibinfo{journal}{\bibinfo{title}{Using structural disorder to enhance the magnetism and spin-polarization in fex si1-x thin films for spintronics}}.
\newblock {\emph{\JournalTitle{Materials Research Express}}} \textbf{\bibinfo{volume}{1}}, \bibinfo{pages}{026102}, \doiprefix\url{10.1088/2053-1591/1/2/026102} (\bibinfo{year}{2014}).

\bibitem{coey1978}
\bibinfo{author}{Coey, J. M.~D.}
\newblock \bibinfo{journal}{\bibinfo{title}{{Amorphous magnetic order}}}.
\newblock {\emph{\JournalTitle{Journal of Applied Physics}}} \textbf{\bibinfo{volume}{49}}, \bibinfo{pages}{1646--1652}, \doiprefix\url{10.1063/1.324880} (\bibinfo{year}{1978}).

\bibitem{Bouma2020}
\bibinfo{author}{Bouma, D.~S.} \emph{et~al.}
\newblock \bibinfo{journal}{\bibinfo{title}{{Itinerant ferromagnetism and intrinsic anomalous Hall effect in amorphous iron-germanium}}}.
\newblock {\emph{\JournalTitle{Physical Review B}}} \textbf{\bibinfo{volume}{101}}, \bibinfo{pages}{014402}, \doiprefix\url{10.1103/PhysRevB.101.014402} (\bibinfo{year}{2020}).
\newblock \eprint{1908.06055}.

\bibitem{terasawa2020}
\bibinfo{author}{Terasawa, A.} \& \bibinfo{author}{Gohda, Y.}
\newblock \bibinfo{title}{First-principles study of exchange coupling constants in nd$_x$fe$_{1-x}$ amorphous alloy} (\bibinfo{year}{2020}).
\newblock \eprint{2008.07735}.

\bibitem{Lindmaa2013}
\bibinfo{author}{Lindmaa, A.}, \bibinfo{author}{Liz\'arraga, R.}, \bibinfo{author}{Holmstr\"om, E.}, \bibinfo{author}{Abrikosov, I.~A.} \& \bibinfo{author}{Alling, B.}
\newblock \bibinfo{journal}{\bibinfo{title}{Exchange interactions in paramagnetic amorphous and disordered crystalline $\mathrm{CrN}$-based systems}}.
\newblock {\emph{\JournalTitle{Phys. Rev. B}}} \textbf{\bibinfo{volume}{88}}, \bibinfo{pages}{054414}, \doiprefix\url{10.1103/PhysRevB.88.054414} (\bibinfo{year}{2013}).

\bibitem{wu2023_amorphous}
\bibinfo{author}{Wu, W.} \emph{et~al.}
\newblock \bibinfo{title}{{Ultra-small topological spin textures with size of 1.3nm at above room temperature in Fe78Si9B13 amorphous alloy}} (\bibinfo{year}{2023}).

\bibitem{Okubo_et_al:2012}
\bibinfo{author}{Okubo, T.}, \bibinfo{author}{Chung, S.} \& \bibinfo{author}{Kawamura, H.}
\newblock \bibinfo{journal}{\bibinfo{title}{Multiple-$q$ states and the skyrmion lattice of the triangular-lattice heisenberg antiferromagnet under magnetic fields}}.
\newblock {\emph{\JournalTitle{Phys. Rev. Lett.}}} \textbf{\bibinfo{volume}{108}}, \bibinfo{pages}{017206}, \doiprefix\url{10.1103/PhysRevLett.108.017206} (\bibinfo{year}{2012}).

\bibitem{Leonov2015}
\bibinfo{author}{Leonov, A.~O.} \& \bibinfo{author}{Mostovoy, M.}
\newblock \bibinfo{journal}{\bibinfo{title}{{Multiply periodic states and isolated skyrmions in an anisotropic frustrated magnet}}}.
\newblock {\emph{\JournalTitle{Nature Communications}}} \textbf{\bibinfo{volume}{6}}, \bibinfo{pages}{8275}, \doiprefix\url{10.1038/ncomms9275} (\bibinfo{year}{2015}).
\newblock \eprint{1501.02757}.

\bibitem{Kurumaji2019}
\bibinfo{author}{Kurumaji, T.} \emph{et~al.}
\newblock \bibinfo{journal}{\bibinfo{title}{{Skyrmion lattice with a giant topological Hall effect in a frustrated triangular-lattice magnet}}}.
\newblock {\emph{\JournalTitle{Science}}} \textbf{\bibinfo{volume}{365}}, \bibinfo{pages}{914--918}, \doiprefix\url{10.1126/science.aau0968} (\bibinfo{year}{2019}).
\newblock \eprint{1805.10719}.

\bibitem{vonMalottki2017}
\bibinfo{author}{von Malottki, S.}, \bibinfo{author}{Dup{\'e}, B.}, \bibinfo{author}{Bessarab, P.~F.}, \bibinfo{author}{Delin, A.} \& \bibinfo{author}{Heinze, S.}
\newblock \bibinfo{journal}{\bibinfo{title}{Enhanced skyrmion stability due to exchange frustration}}.
\newblock {\emph{\JournalTitle{Scientific Reports}}} \textbf{\bibinfo{volume}{7}}, \bibinfo{pages}{12299} (\bibinfo{year}{2017}).

\bibitem{Montoya2017}
\bibinfo{author}{Montoya, S.~A.} \emph{et~al.}
\newblock \bibinfo{journal}{\bibinfo{title}{Tailoring magnetic energies to form dipole skyrmions and skyrmion lattices}}.
\newblock {\emph{\JournalTitle{Phys. Rev. B}}} \textbf{\bibinfo{volume}{95}}, \bibinfo{pages}{024415}, \doiprefix\url{10.1103/PhysRevB.95.024415} (\bibinfo{year}{2017}).

\bibitem{Kresse1996}
\bibinfo{author}{Kresse, G.} \& \bibinfo{author}{Furthm{\"{u}}ller, J.}
\newblock \bibinfo{journal}{\bibinfo{title}{{Efficient iterative schemes for {\textless}i{\textgreater}ab initio{\textless}/i{\textgreater} total-energy calculations using a plane-wave basis set}}}.
\newblock {\emph{\JournalTitle{Physical Review B}}} \textbf{\bibinfo{volume}{54}}, \bibinfo{pages}{11169--11186}, \doiprefix\url{10.1103/PhysRevB.54.11169} (\bibinfo{year}{1996}).

\bibitem{Kresse1999}
\bibinfo{author}{Kresse, G.} \& \bibinfo{author}{Joubert, D.}
\newblock \bibinfo{journal}{\bibinfo{title}{{From ultrasoft pseudopotentials to the projector augmented-wave method}}}.
\newblock {\emph{\JournalTitle{Physical Review B}}} \textbf{\bibinfo{volume}{59}}, \bibinfo{pages}{1758--1775}, \doiprefix\url{10.1103/PhysRevB.59.1758} (\bibinfo{year}{1999}).

\bibitem{Blochl1994}
\bibinfo{author}{Bl{\"{o}}chl, P.~E.}
\newblock \bibinfo{journal}{\bibinfo{title}{{Projector augmented-wave method}}}.
\newblock {\emph{\JournalTitle{Physical Review B}}} \textbf{\bibinfo{volume}{50}}, \bibinfo{pages}{17953--17979}, \doiprefix\url{10.1103/PhysRevB.50.17953} (\bibinfo{year}{1994}).

\bibitem{Perdew1992}
\bibinfo{author}{Perdew, J.~P.} \& \bibinfo{author}{Wang, Y.}
\newblock \bibinfo{journal}{\bibinfo{title}{{Accurate and simple analytic representation of the electron-gas correlation energy}}}.
\newblock {\emph{\JournalTitle{Physical Review B}}} \textbf{\bibinfo{volume}{45}}, \bibinfo{pages}{13244--13249}, \doiprefix\url{10.1103/PhysRevB.45.13244} (\bibinfo{year}{1992}).

\bibitem{Xiang2013}
\bibinfo{author}{Xiang, H.}, \bibinfo{author}{Lee, C.}, \bibinfo{author}{Koo, H.-J.}, \bibinfo{author}{Gong, X.} \& \bibinfo{author}{Whangbo, M.-H.}
\newblock \bibinfo{journal}{\bibinfo{title}{{Magnetic properties and energy-mapping analysis}}}.
\newblock {\emph{\JournalTitle{Dalton Trans.}}} \textbf{\bibinfo{volume}{42}}, \bibinfo{pages}{823--853}, \doiprefix\url{10.1039/C2DT31662E} (\bibinfo{year}{2013}).

\bibitem{Xu2019}
\bibinfo{author}{Xu, C.}, \bibinfo{author}{Xu, B.}, \bibinfo{author}{Dup{\'{e}}, B.} \& \bibinfo{author}{Bellaiche, L.}
\newblock \bibinfo{journal}{\bibinfo{title}{{Magnetic interactions in BiFeO 3 : A first-principles study}}}.
\newblock {\emph{\JournalTitle{Physical Review B}}} \textbf{\bibinfo{volume}{99}}, \bibinfo{pages}{104420}, \doiprefix\url{10.1103/PhysRevB.99.104420} (\bibinfo{year}{2019}).

\bibitem{Bayaraa2021a}
\bibinfo{author}{Bayaraa, T.}, \bibinfo{author}{Xu, C.} \& \bibinfo{author}{Bellaiche, L.}
\newblock \bibinfo{journal}{\bibinfo{title}{{Magnetization Compensation Temperature and Frustration-Induced Topological Defects in Ferrimagnetic Antiperovskite $\text{Mn}_4\text{N}$}}}.
\newblock {\emph{\JournalTitle{Physical Review Letters}}} \textbf{\bibinfo{volume}{127}}, \bibinfo{pages}{217204}, \doiprefix\url{10.1103/PhysRevLett.127.217204} (\bibinfo{year}{2021}).

\bibitem{tb2j}
\bibinfo{author}{He, X.}, \bibinfo{author}{Helbig, N.}, \bibinfo{author}{Verstraete, M.~J.} \& \bibinfo{author}{Bousquet, E.}
\newblock \bibinfo{journal}{\bibinfo{title}{{TB2J: A python package for computing magnetic interaction parameters}}}.
\newblock {\emph{\JournalTitle{Computer Physics Communications}}} \textbf{\bibinfo{volume}{264}}, \bibinfo{pages}{107938}, \doiprefix\url{10.1016/j.cpc.2021.107938} (\bibinfo{year}{2021}).
\newblock \eprint{2009.01910}.

\bibitem{Miyatake1986}
\bibinfo{author}{Miyatake, Y.}, \bibinfo{author}{Yamamoto, M.}, \bibinfo{author}{Kim, J.~J.}, \bibinfo{author}{Toyonaga, M.} \& \bibinfo{author}{Nagai, O.}
\newblock \bibinfo{journal}{\bibinfo{title}{{On the implementation of the 'heat bath' algorithms for Monte Carlo simulations of classical Heisenberg spin systems}}}.
\newblock {\emph{\JournalTitle{Journal of Physics C: Solid State Physics}}} \textbf{\bibinfo{volume}{19}}, \bibinfo{pages}{2539--2546}, \doiprefix\url{10.1088/0022-3719/19/14/020} (\bibinfo{year}{1986}).

\bibitem{PASP}
\bibinfo{author}{Lou, F.} \emph{et~al.}
\newblock \bibinfo{journal}{\bibinfo{title}{{PASP: Property analysis and simulation package for materials}}}.
\newblock {\emph{\JournalTitle{The Journal of Chemical Physics}}} \textbf{\bibinfo{volume}{154}}, \bibinfo{pages}{114103}, \doiprefix\url{10.1063/5.0043703} (\bibinfo{year}{2021}).

\bibitem{Berg1981}
\bibinfo{author}{Berg, B.} \& \bibinfo{author}{L{\"{u}}scher, M.}
\newblock \bibinfo{journal}{\bibinfo{title}{{Definition and statistical distributions of a topological number in the lattice O(3) $\sigma$-model}}}.
\newblock {\emph{\JournalTitle{Nuclear Physics, Section B}}} \textbf{\bibinfo{volume}{190}}, \bibinfo{pages}{412--424}, \doiprefix\url{10.1016/0550-3213(81)90568-X} (\bibinfo{year}{1981}).

\bibitem{zenodo_doi}
\bibinfo{author}{Bayaraa, T.}
\newblock \bibinfo{title}{Amorphous fege paper data: v0}, \doiprefix\url{10.5281/zenodo.19053623} (\bibinfo{year}{2026}).

\end{thebibliography}

\end{document}


\date{\today}

\begin{abstract}
This Supplemental Material provides extended results, methodological details, and statistical analyses that complement the main manuscript on disorder-driven topological spin textures in amorphous FeGe. While the main text focuses on the emergent disorder-driven magnetic behavior of amorphous FeGe and its connection to topological spin textures, here we present (i) detailed benchmarks of our first-principles-based methodology against the crystalline B20 FeGe reference, (ii) validation of the amorphous structure generation protocol, including a comparison of different melt-quench schemes, (iii) a comprehensive statistical characterization of the local structural, chemical, and magnetic environments of Fe ions in the amorphous phase, and (iv) a detailed analysis of the bond-resolved magnetic interactions—exchange coupling, Dzyaloshinskii--Moriya interaction (DMI), and single-ion anisotropy (SIA)—that underpin the spin Hamiltonian used in the manuscript. Together, these results substantiate the central finding that structural and chemical disorder in amorphous FeGe gives rise to broad distributions of magnetic interaction parameters that cannot be captured by conventional distance-only models, and that this heterogeneity is essential for stabilizing the topological spin textures reported in the main text.
\end{abstract}

\maketitle
\beginsupplement

\section*{Organization of the Supplemental Material}

The Supplemental Material is organized into five Supplementary Notes:

\noindent\textbf{Supplementary Note 1: Generation and validation of crystalline B20 and amorphous FeGe structures.} We detail the DFT calculation parameters, including the choice of Hubbard $U$ correction and exchange--correlation functional, and benchmark our results for crystalline B20 FeGe against experiment (Tables S1 and S2, Fig.~S1). We then describe the \textit{ab initio} melt-quench protocol used to generate amorphous snapshots, compare PBE and LDA functionals for the amorphous phase (Table S3), and evaluate three different quenching schemes—stepwise quenching, rapid quenching, and cooling—based on their resulting atomic number densities and radial distribution functions (Fig.~S3). Simulated spin textures from Monte Carlo simulations of the crystalline reference are also shown (Fig.~S2).

\noindent\textbf{Supplementary Note 2: Distribution of local structural and magnetic environments.} We perform a statistical analysis of 71 amorphous structures, characterizing the distributions of Fe magnetic moments, nearest-neighbor counts (Fe, Ge, and total), and average bond distances (Fig.~S4). We identify the most common local neighbor patterns surrounding Fe atoms (Fig.~S5) and investigate correlations between the magnitude and sign of the local magnetic moments and various structural descriptors, including bond angles (Fig.~S8) and local structural motifs visualized using a barcode representation (Fig.~S9). Finally, we demonstrate that the sign of the Fe magnetic moments is governed primarily by the local mean field generated by neighboring spins (Figs.~S6, S7, and S10).

\noindent\textbf{Supplementary Note 3: Bond-resolved magnetic interactions in amorphous FeGe.} We present the full distributions of exchange couplings $J_{ij}$, DMI vectors, and SIA tensors extracted for the amorphous structures (Figs.~S11 and S12), and show that these interactions depend jointly on interatomic distance \textit{and} local moment amplitude—a departure from conventional distance-only Bethe--Slater models. We also examine the possibility of superexchange via Fe--Ge--Fe pathways (Fig.~S13) and analyze how local coordination environments influence the single-ion anisotropy using the barcode method (Fig.~S14).

\noindent\textbf{Supplementary Note 4: Construction of the spin Hamiltonian.} We describe the decomposition of the $3\times3$ exchange matrices into symmetric (collinear exchange) and antisymmetric (DMI) parts, define the relation between the antisymmetric component and the DMI vector $\mathbf{D}$, and outline the four-state energy mapping procedure used to extract all magnetic interaction parameters reported in the manuscript.

\noindent\textbf{Supplementary Note 5: Barcode method.} We describe in detail the barcode representation developed to visualize and analyze the local atomic environments around Fe ions in the amorphous structures, and discuss its advantages over standard structural fingerprint methods for capturing the geometric heterogeneity characteristic of amorphous systems.

\section*{Supplementary Note 1: Generation and validation of crystalline B20 and amorphous FeGe structures}

\subsection*{DFT calculation parameters}

All first-principles calculations were carried out within the framework of Density Functional Theory (DFT) as implemented in Vienna \textit{Ab-Initio} Software Package (VASP) \cite{Kresse1999} using the projector augmented-wave potentials \cite{Blochl1994}. For the crystalline case, we used the generalized gradient approximation, with the Perdew-Burke-Ernzerhof exchange-correlation functional \cite{Liechtenstein1995,GGA}, and an effective Hubbard U parameter of 1.5 eV for the localized 4p electrons of the Ge ions.  Typically, the Hubbard U parameter is introduced on localized manifolds to account for the underlocalization inherent in the mean-field treatment in DFT and is often benchmarked against experimental observations such as lattice parameters and magnetic moments. However, since FeGe is a metal, we do not expect there to be a need for significant corrections of the Fe-d manifold. Interestingly, as was previously reported, a better agreement with experimental lattice parameters for B20 FeGe is obtained with a U parameter on the Ge-4p as discussed in Ref. \cite{grytsiuk2019}.

Table S1 reports the calculated lattice parameter $a$ and magnetic moment of Fe ions for different Hubbard $U$ parameters applied using the Dudarev approximation on the 3d and 2p orbitals of Fe and Ge ions, respectively. Experimental results \cite{lebech1989} are also shown to provide a comparison. We find that larger U values on Fe-d lead to a much larger magnetic moment than those reported in experiment, for instance, a U value of 1.5 eV gives 2.27 $\mu_B$, which is more than double the experimental result of 1 $\mu_B$. On the other hand, introducing a Hubbard U on Ge-p brings the lattice parameters into close agreement with experiments without enhancing the magnitude of the magnetic moments of Fe ions. For example, for a U of 1.5 eV on Ge-p, we calculate the lattice parameter and magnetic moment to be 4.69 \AA~ and 1.19 $\mu_B$, which is in good agreement with the experimental result of 4.69 \AA~ and 1.00 $\mu_B$ at 10 K~\cite{Spencer2018}. A similar finding was reported previously in Ref.~\cite{grytsiuk2019} that applying the Hubbard-U correction (U = 1.5 eV and J = 0.5 eV) to Ge p-orbitals was found to better reproduce the experimentally measured magnetic moments of Fe. However, they found that such Hubbard U treatment increases the length of the spin spiral found in FeGe. Thus, we chose to apply Hubbard U of 1.5 eV on the 4p orbitals of Ge ions throughout the study of crystalline FeGe structure in the manuscript and hereafter.

For the amorphous FeGe case, neither the Hubbard U nor the Wigner--Seitz radius was used or specified in our calculations. This is due to the fact that we found diverse local structural, chemical, and magnetic environments which can not be individually assigned a Hubbard U or Wigner--Seitz radius in VASP.

We used an energy cutoff of 860 eV and a Monkhorst-Pack k-point mesh density of $13\times13\times13$ and $\Gamma$-point only sampling for the Brillouin Zone (BZ) of the crystalline and amorphous cases, respectively. All structural relaxations, both crystalline and amorphous, were performed until the Hellmann-Feynman force on each atom was less than 0.001 eV/\AA\ and included spin-orbit coupling (SOC) self-consistently as implemented in VASP.

\begin{table}[htbp]
    \centering
    \caption{Lattice parameter and magnetic moment of Fe using different Hubbard U parameters on 3d and 2p orbitals of Fe and Ge for the crystalline B20 phase, respectively.}
    \label{tab:table1}
    \resizebox{0.45\textwidth}{!}{%
    \begin{tabular}{@{}cSS@{}}
        \toprule
        & {Lattice parameter (\AA)} & {Magnetic moment (\(\mu_B\))} \\ 
        \midrule
        U on Fe (eV) & & \\
        0.5 & 4.67 & 1.30 \\
        1.0 & 4.70 & 1.73 \\
        1.5 & 4.77 & 2.27 \\
        U on Ge (eV) & & \\
        1.0 & 4.68 & 1.18 \\
        1.5 & 4.69 & 1.19 \\
        2.0 & 4.70 & 1.19 \\
        \midrule
        Without U & 4.66 & 1.17 \\
        Exp. results \cite{lebech1989} & 4.69 & 1.00 \\
        Other theory \cite{grytsiuk2019} & 4.67 & 1.16 \\
        \bottomrule
    \end{tabular}%
    }
\end{table}

\begin{table}[htbp]
\centering
\caption{Comparison of different functionals on lattice parameters and magnetic moments for the crystalline B20 phase}
\label{tab:lattice_magnetic_moments}
\resizebox{0.5\textwidth}{!}{
\begin{tabular}{
  l
  c
  S[table-format=1.2] 
  S[table-format=1.2] 
}
\toprule
 & {Method} & {Lattice parameter (\AA)} & {Magnetic moment ($\mu_B$)} \\
\midrule
Without U & PBE & 4.66 & 1.17 \\
 & LDA & 4.55 & 1.11 \\
With U of & PBE & 4.69 & 1.18 \\
1.5 eV on Ge & LDA & 4.58 & 1.10 \\
\midrule
Exp. results \cite{lebech1989} & & 4.69 & 1.00 \\
\bottomrule
\end{tabular}
}
\end{table}

\begin{figure*}
\includegraphics[width=0.8\textwidth]{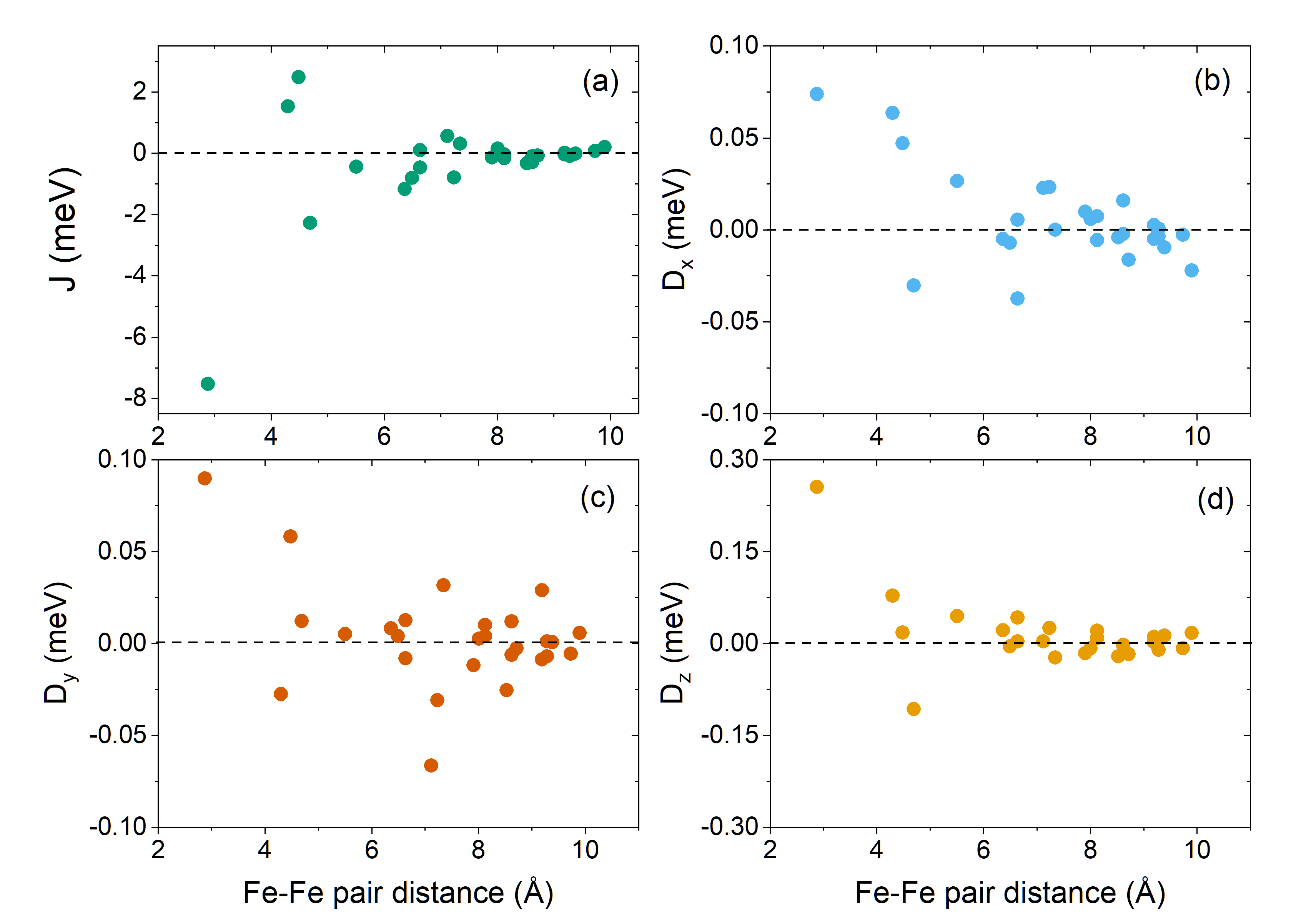}
\caption{Calculated magnetic exchange coupling (a) and three components of the DMI (b-d) of crystalline FeGe are shown as a function of the Fe--Fe pair distances.}
\end{figure*}

\subsection*{Crystalline B20 FeGe}

Fig.~S1 plots the calculated magnetic exchange coupling coefficient (a) and three components of DMI (b-d) as a function of the Fe--Fe pair distance for B20 FeGe. All parameters are calculated using the Green's function method \cite{tb2j} after obtaining Wannier functions. The \textit{ab initio} DFT Bloch wavefunction was projected onto highly symmetric atomic-orbital-like Wannier functions \cite{Wannier90} as implemented in the VASP code. We included the outermost d orbitals for Fe and p orbitals for Ge to cover the full band overlap from the \textit{ab initio} and Wannier functions. We find the 1NN interaction to be FM, followed by AFM 2NN and 3NN interactions. We find that longer-range interactions beyond these are mostly FM (see Fig.~S1(a)). These competing magnetic interactions at different ranges result in geometrically frustrated spin interactions. Moreover, our results of J and DMI match well with previous studies \cite{grytsiuk2019,grytsiuk2021,borisov2021}.

\begin{figure*}
\includegraphics[width=0.8\textwidth]{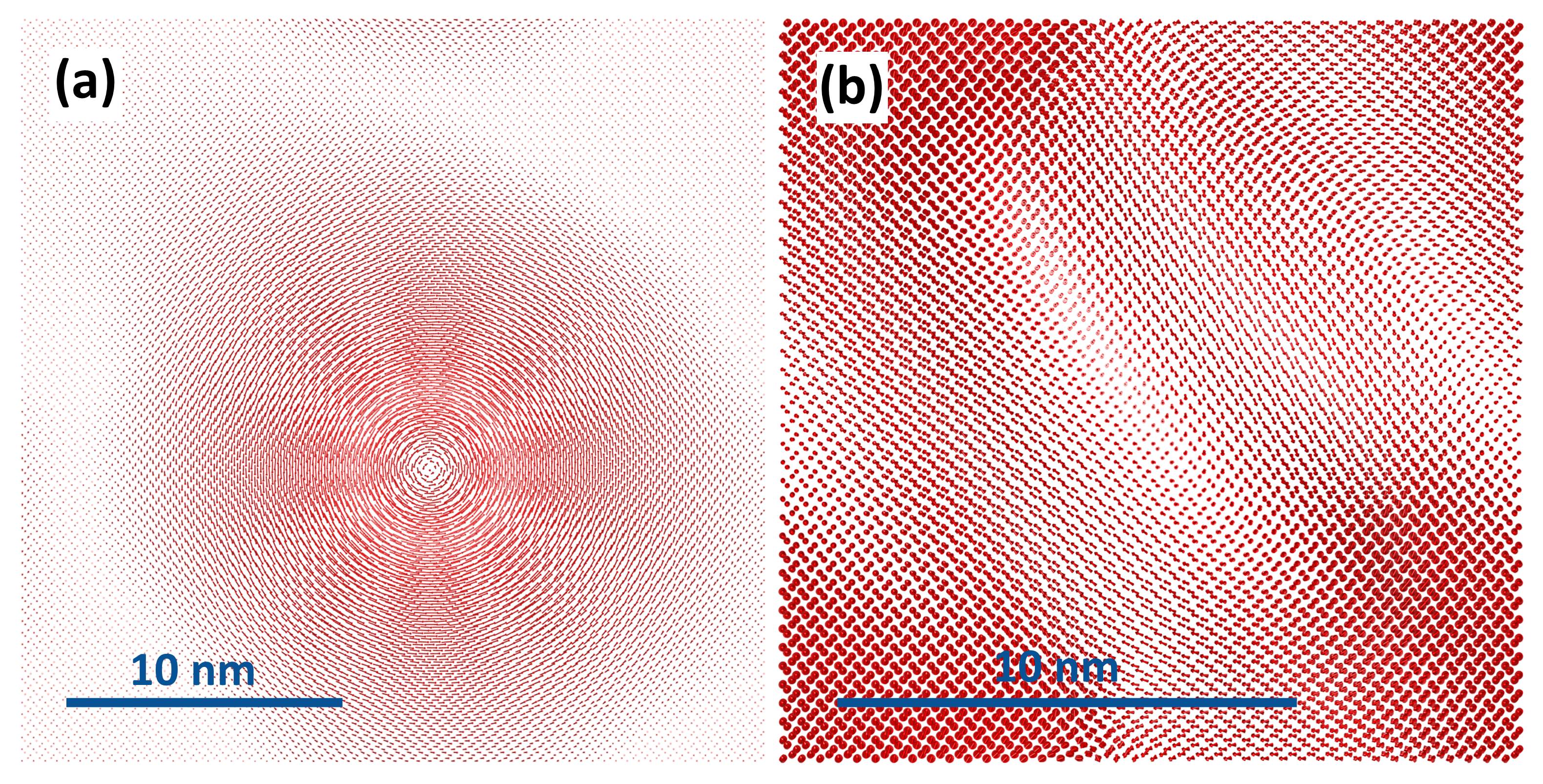}
\caption{Simulated spin textures of crystalline FeGe (a) and under an external magnetic field (b) are shown. To clearly show the skyrmion texture, the studied supercell is extended by 2 times along the in-plane directions.}
\end{figure*}

Using the extracted magnetic parameters and spin Hamiltonian of Eqs. (1) and (2) from the manuscript, we ran Parallel Tempering Monte Carlo (MC) simulations \cite{Miyatake1986,PASP} with a conjugate gradient (CG) method \cite{Hestenes1952}. We used supercells of different sizes for the crystalline structures, 64$\times$64$\times$2, 128$\times$128$\times$4, 24$\times$24$\times$24, and 40$\times$40$\times$40, and at each temperature, 160,000 Monte Carlo sweeps were performed. Note that a 40$\times$40$\times$40 supercell is the largest supercell we could run with MC simulation due to the computational limitations. The predicted Curie temperature, $T_C$, was 235 K for 40$\times$40$\times$40 supercell, which was lower than the experimental value of $\sim$280 K \cite{Spencer2018}, and we found $T_C$ increasing with larger supercell sizes, suggesting that reaching the experimental $T_C$ of 280 K is possible if one could consider a large enough supercell in their simulations. In addition, we use the Curie--Weiss law in the mean field approximation and calculated $T_C$ including the exchange coupling coefficients until 6th nearest neighbor, $T_C = S(S+1)/(3 \times k_B) \times \sum^{6NN}_{ij} J_{ij}$, we find that $T_C$ is 281 K which is in good agreement with experimental value of $T_C$. Furthermore, we found different spin textures from the MC simulations, some of which were in good comparison with the experimental results \cite{turgut2017,zheng2018,zhang2017,kovacs2016,yu2011}. Note that we were not able to get clear helical spin spiral textures due to the finite-size effect and computational limitations. Fig.~S2 shows the simulated spin textures of skyrmions (a) from a 64$\times$64$\times$2 supercell and chiral spin texture (b) from a 40$\times$40$\times$40 supercell in B20 FeGe. Fig.~S2(b) shows a sign of chiral spin texture at the center, which hints toward finding a clear helical spin spiral phase over a longer period if larger supercells can be considered in the MC simulations. Thus, finding an experimentally observed helical phase in the B20 FeGe structure is possible if one can overcome the finite-size effect or computational limitations in MC simulations.

\subsection*{Ab initio melt-quench protocol}

After obtaining an optimized crystalline structure, we used \textit{ab-initio} molecular dynamics (AIMD) simulations with the NVT ensemble as implemented in VASP to generate amorphous snapshots. Such "melt-quench" methodology was previously demonstrated on several systems \cite{corbae2023,harrelson2021,cheng2020,sivonxay2020,sivonxay2022}, including Fe$_x$Ge$_{1-x}$ \cite{Zoey2025,Bouma2020,karel2020}. Atoms were first randomly distributed in a cubic simulation cell using the Packmol package \cite{packmol}. The supercell size was chosen to contain 96 atoms to balance a large enough size to obtain representative amorphous environments with a reasonable runtime of AIMD simulations and the number of unique Fe--Fe pairs to be studied for the extraction of the magnetic exchange parameters (note that the crystalline unit cell has 8 atoms). All our amorphous calculations were for a Fe$_{0.5}$Ge$_{0.5}$ stoichiometry. To obtain a stable liquid phase, the pressure was equilibrated through a series of AIMD simulations at 3000 K, with rescaling of the unit cell between each AIMD simulation until an internal pressure of 0 bar. Afterward, we performed a 10 ps simulation after the energy equilibration process, collecting ten snapshots of the melted structures at regular intervals.

\subsection*{Comparison between different functionals}

We also checked the effect of different exchange-correlation functionals, such as Local-density approximations (LDA)\cite{LDA} and Perdew-Burke-Ernzerhof (PBE)\cite{GGA}, on the lattice parameters and magnetic moments. For the crystalline B20 phase, the results are shown in Table S2, and in general, both functionals underestimate the lattice parameter but overestimate the magnetic moment of the experimental values \cite{lebech1989}, and it is in good agreement with the previous study \cite{grytsiuk2019}.

The calculated average magnetic moments of Fe ions in five amorphous snapshots acquired from AIMD calculations using LDA\cite{LDA} and PBE\cite{GGA} exchange-correlation functionals are shown in Table S3. The average magnetic moments of Fe ions were found to be about twice the experimental value of 0.75 $\mu_B$ \cite{streubel2021} when the PBE exchange-correlation functional was used. On the other hand, when using the LDA exchange-correlation functional, the average magnetic moments of Fe ions are calculated to be smaller than the ones generated by PBE but much closer to the experimental value. For instance, one of the snapshots was calculated to have an average magnetic moment of 0.79 $\mu_B$, which is in good agreement with the experimental value of 0.75 $\mu_B$ \cite{streubel2021}. To understand the difference in magnetic moments when using PBE and LDA functionals, we calculated the average Fe--Ge and Fe--Fe bond lengths. The average Fe--Ge bond lengths were found to be 2.59\AA~ and 2.54\AA~ for PBE and LDA functionals, respectively. The average Fe--Fe bond lengths were found to be 2.71\AA~ and 2.66\AA~ for PBE and LDA functionals, respectively. Thus, when the PBE functional is used, Fe \textit{d}-orbitals get more localized, and larger magnetic moments are predicted from the calculations with respect to the case when the LDA functional is used. In addition, we found a 5-6\% increase in the volume of amorphous supercells when the PBE functional is used in comparison to when the LDA functional is used in the calculations, which shows that the PBE functional does not treat the amorphous structures as tightly packed as the LDA functional does, and amorphous metals are characterized to have random-dense packing. Note that in the crystalline case, both the LDA and PBE functionals overestimate the experimental magnetic moment, and in fact, PBE predicts a larger magnetic moment than the one from the LDA functional. In addition, Ref. \cite{grytsiuk2019} reports that the helical spin spiral period is overestimated by a factor of two when using the parameters obtained by both PBE and LDA functionals. Moreover, the LDA functional was generally found to work well for amorphous silicates and oxides \cite{karki2015}. Based on these results, a good choice would be to use the LDA exchange-correlation functional for the study of amorphous structures. However, as reported here and in the manuscript, one needs to sample enough spin configurations to obtain amorphous magnetic structures, and a larger volume brings the atomic number density of our structures closer to the experimental value; thus, we used the PBE functional for both the crystalline and amorphous FeGe systems. We report the atomic number densities of our amorphous snapshots obtained using PBE and LDA functionals in Table S3, which shows that amorphous snapshots obtained with LDA underestimate the experimental volume by about 13 to 16\%.

\begin{table}[htbp]
\centering
\caption{Calculated average magnetic moments of Fe ions and atomic number density in amorphous structures}
\label{tab:table2}
\begin{tabular}{
  c
  S[table-format=1.2] 
  S[table-format=1.2]
  S[table-format=1.2] 
  S[table-format=1.2]
}
\toprule
{Amorphous snapshots} & \multicolumn{2}{c}{Magnetic moment ($\mu_B$)} & \multicolumn{2}{c}{Atomic number density (10$^{22}$~cm$^{-3}$)} \\
& {PBE} & {LDA} & {PBE} & {LDA}\\
\midrule
1 & 1.41 & 0.66 & 7.15 & 7.90\\
2 & 1.07 & 0.52 & 7.13 & 7.89\\
3 & 1.49 & 0.57 & 6.93 & 7.60\\
4 & 1.55 & 0.57 & 6.98 & 7.70\\
5 & 1.53 & 0.79 & 7.04 & 7.73\\
\midrule
Exp. results \cite{streubel2021,Bouma2020} & \multicolumn{2}{c}{0.75} & \multicolumn{2}{c}{6.6}\\
\bottomrule
\end{tabular}
\end{table}

\subsection*{Comparison between different quenching schemes}

\begin{figure*}
\includegraphics[width=0.8\textwidth]{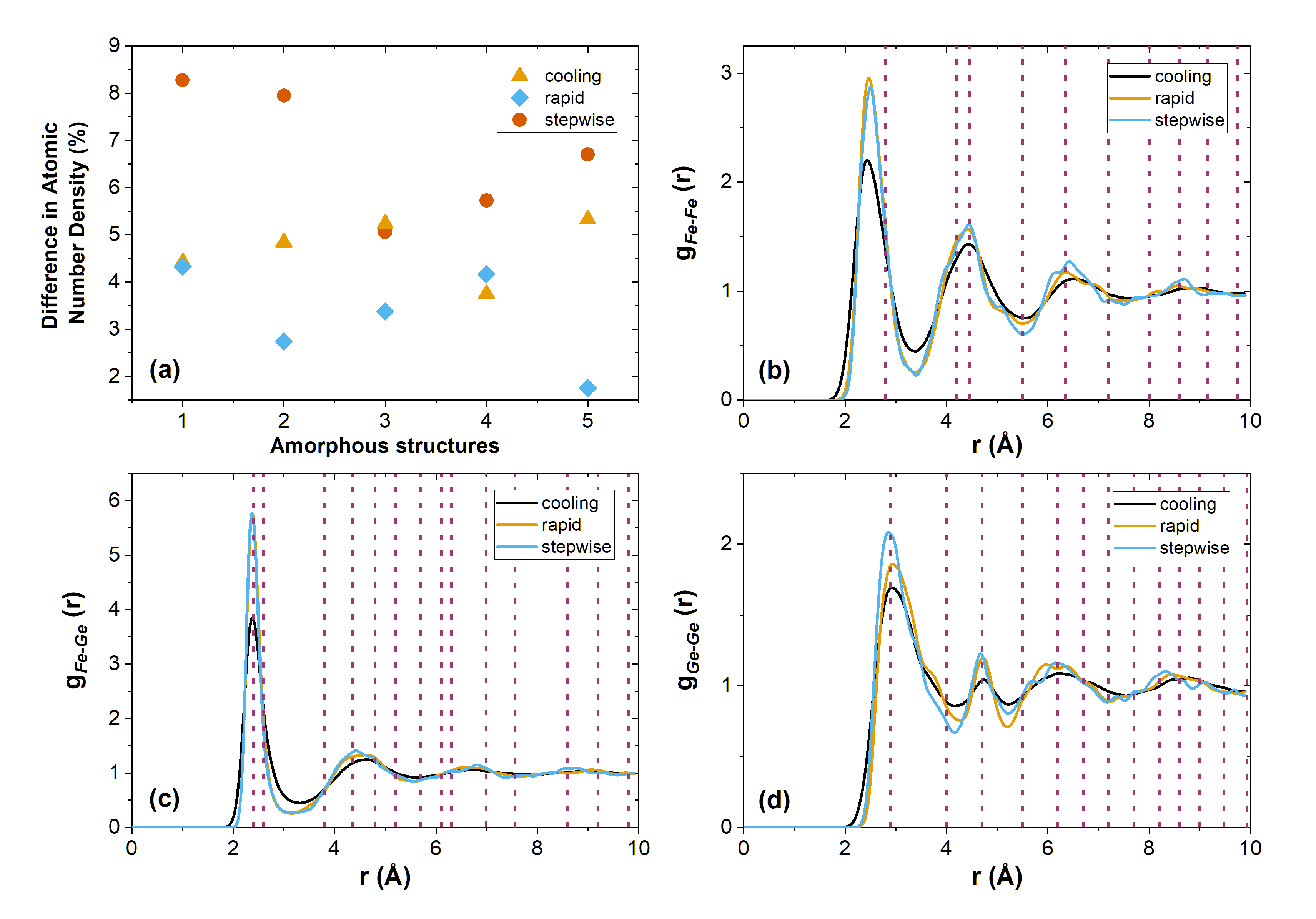}
\caption{Comparison between three different quenching schemes. Panel (a) shows the difference in the atomic number density compared to the experimental value in Ref.\cite{Bouma2020} for 5 different structures relaxed with DFT after the quenching from all schemes. Panels (b-d) report the radial distribution functions between Fe--Fe ions, Fe--Ge ions, and Ge--Ge ions for amorphous structures generated by three different schemes. Radial distribution functions are averaged between 5 structures and taken into account for the last 500 steps from the \textit{ab-initio} molecular dynamics simulations. Dashed purple lines represent neighbor distances from crystalline counterpart.}
\end{figure*}

As indicated in the manuscript and reported in the previous work~\cite{Bouma2020}, there is up to a 10\% difference in the atomic number density between theory and experiment for different chemical compositions of amorphous Fe$_x$Ge$_{1-x}$. Thus, in order to overcome this issue, we investigated and applied three different quenching schemes in our work: stepwise quenching, rapid quenching, and cooling. Stepwise quenching uses a stepped cooling scheme with 400 fs quenching of 500 K, followed by 1 ps isothermal steps. In the rapid quenching scheme, we directly changed the temperature of the system from 3000 K to 200 K and equilibrated for 10 ps (note that under a temperature below 200 K, the system becomes unstable), and in the cooling scheme, we gradually decreased the temperature from 3000 K to 10 K for 10 ps. For the stepwise quenching scheme, the volume of the cell was optimized by checking the external pressure during the isothermal steps, and for both the rapid quenching and cooling schemes, the volume was kept frozen to a volume with experimental atomic number density. These three different schemes were used on 5 different liquid snapshots, randomly chosen from 10 snapshots indicated in the manuscript, and then optimized at 0 K using VASP with the higher precision parameters mentioned previously. Fig.~S3(a) shows the difference in the atomic number densities of these 15 amorphous structures with respect to the experimental value of 6.6 $\times$ 10$^{22}$~cm$^{-3}$ ~\cite{Bouma2020}. Surprisingly, our results show that structures obtained with the rapid quenching scheme give much more experimentally accurate ones, compared to the structures obtained through the other two schemes. On average, amorphous structures from stepwise, cooling, rapid quenching schemes have 6.7\%, 4.7\%, and 3.3\% larger atomic number density compared to the experimental value, respectively. Although it requires further and careful analysis, we hypothesize that, with a rapid quenching scheme, one has a greater tendency to retain local motifs from the liquid phase. In the other two schemes, the system has freedom and time to relax and change into different local motifs. Next, let us look at the comparison between radial distribution functions (RDFs) of amorphous structures generated by the three quenching schemes. Fig.~S3(b-d) shows the averaged radial distribution functions between Fe--Fe, Fe--Ge, and Ge--Ge pairs of amorphous structures obtained by the three quenching schemes. One can see that all structures have an amorphous pair correlation nature and have similar function shapes with the same peak locations. In addition, our results show that RDFs from the cooling scheme have broader and more evenly distributed shapes compared to the other two quenching schemes. This can be understood with the fact that in the cooling scheme, the temperature is gradually decreased, hence there is more time and freedom for atoms to arrange into different structure motifs when temperature changes in the simulations. Now, let us do some comparison between nearest neighbor (NN) distances between amorphous and crystalline FeGe. Fig.~S3(b) shows that the first NN (1NN) distance between Fe atoms is shorter in amorphous structures compared to the crystalline counterpart; however, the second NN (2NN) has a similar distance. In addition, we do not see a peak in amorphous RDF where there is 4NN at 5.5\AA~ in crystalline FeGe, but there is a peak in amorphous RDF around 6.4\AA~ overlapping with crystalline 5NN distance. For the RDF between Fe and Ge atoms, Fig.~S3(c) shows that the 1NN distance stays fairly similar between amorphous and crystalline systems, but most of the higher crystalline NN distances are combined/distributed into several broad peaks, as shown in the amorphous RDFs. Lastly, Fig.~S3(d) shows us that RDF between Ge atoms has a peak in 1NN distance at a distance similar to the crystalline counterpart, but no peaks near the 2NN and 4NN distances of the crystalline structure. These results show that amorphous structures have an overall different chemical bonding environment compared to the crystalline counterpart. Note that these RDFs are from \textit{ab-initio} molecular dynamics simulations and do not include magnetism.      

\section*{Supplementary Note 2: Distribution of local structural and magnetic environments}

\subsection*{Distribution of magnetic moments, bond lengths, nearest neighbors, average neighbor distances, and common neighbors}

We performed statistical analysis on the 71 structures reported in Fig.~1(b) of the manuscript to have an in-depth understanding of the various local environments in amorphous FeGe and show the results in Fig.~S4. Panel (a) shows the distribution of magnetic moments in these structures, ranging from -2.6 $\mu_B$ to 2.6 $\mu_B$, with the majority of them being around ±2 $\mu_B$. Note that negative magnetic moments are with respect to the default global magnetic easy axis in VASP. This shows that in amorphous magnets, one can expect to have complex magnetic interactions and Fe atoms to be in high- and low- spin states in one structure. Interestingly, some structures were found to have two ferromagnetic domains with opposite sign moments and an antiferromagnetic domain wall between them, leading to a smaller magnetic moment per Fe (the total magnetic moment divided by the number of Fe atoms). Panels (b-d) show the distribution of the number of Fe, Ge, and total neighbors around each Fe atom within the sphere of 3.5\AA~, respectively. 3.5\AA~radius cutoff was chosen from analyzing RDFs in Fig.~S3(b-d) where the first peak ends for g$_{Fe--Fe}$ and g$_{Fe--Ge}$. Our analysis shows that the expected number of neighbors increases for amorphous structures compared to the crystalline counterpart. Panel (b) reveals that Fe atoms can have as few as 2 to as many as 16 Fe neighbors and, on average, have 7-8 Fe neighbors, an increase with respect to the crystalline value of 6. Similarly, panel (c) shows that the number of Ge neighbors to Fe atoms increased to 8 from 7 in the crystalline counterpart, but can vary between 4 and 13. In total, our analysis reveals that Fe atoms in amorphous structures have more neighbors compared to the crystalline counterpart, with a mean value of 16 total neighbors. Panels (e) and (f) illustrate the distribution of average Fe--Fe and Fe--Ge distances, respectively. Note that this distribution is not analogous to RDFs reported in Fig.~S3, and here we show the distribution of average distance between Fe--Fe and Fe--Ge pairs within each 3.5\AA~radius sphere, in other words, this shows the average distance between the center Fe with its neighboring Fe and Ge atoms, and plots the distribution of these average distances across all structures. Thus, in order to correctly take into account the effective magnetic field surrounding Fe atoms, one needs to take into account pairs within at least 3\AA~, and it can be seen from Fig.~4 of the manuscript and Fig.~S11 that within this range, J parameters are not negligible.

\begin{figure*}
\includegraphics[width=0.8\textwidth]{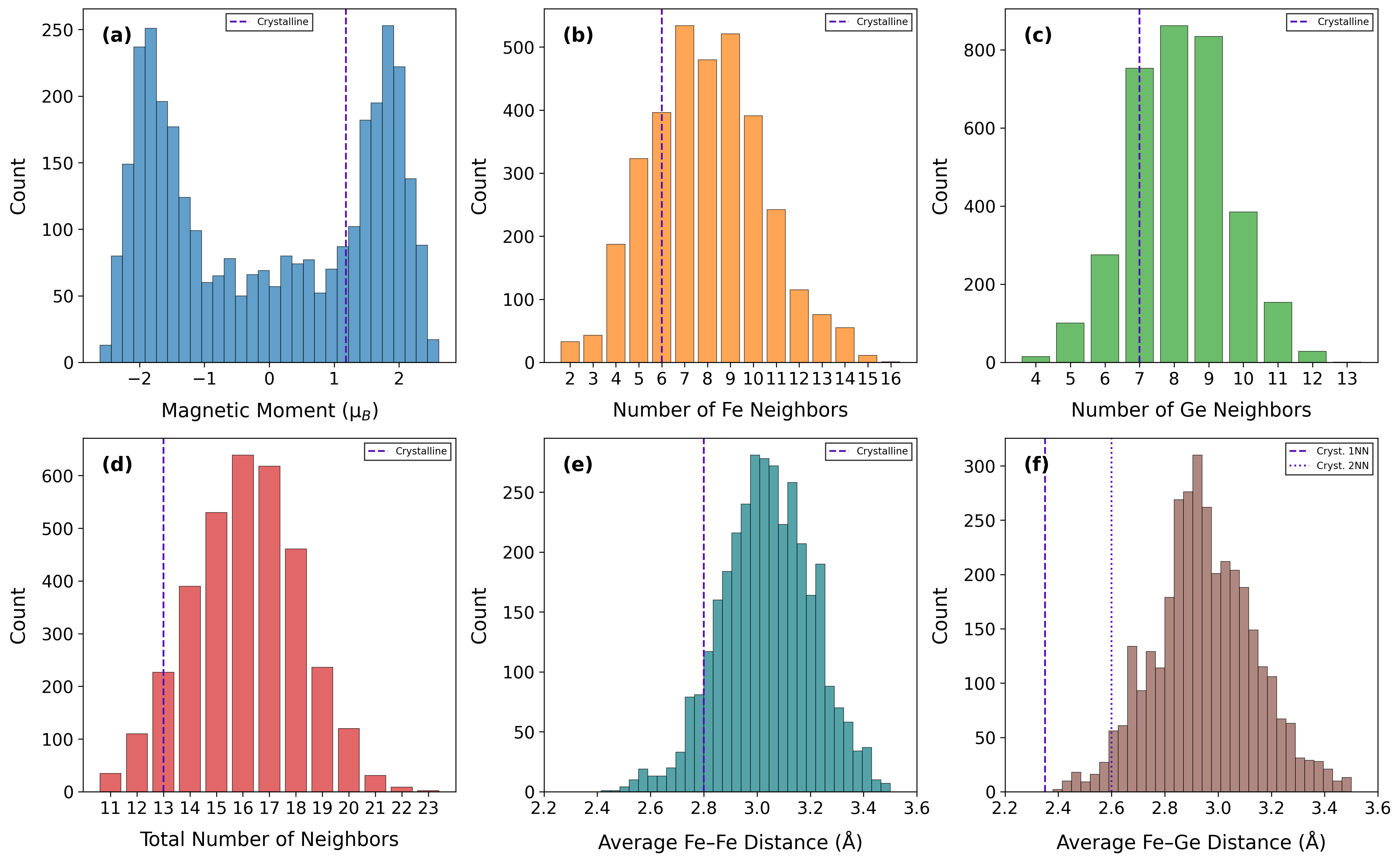}
\caption{Distribution of the magnetic moment (a), Fe neighbors (b), Ge neighbors (c), total neighbors (d), average Fe--Fe distance (e), and average Fe--Ge distance for the structures shown in Fig.~1(b) of the manuscript. Analysis was done within the sphere of 3.5\AA~radius. Vertical dashed lines represent references from B20 crystal FeGe structure.}
\end{figure*}

\begin{figure*}
\includegraphics[width=0.8\textwidth]{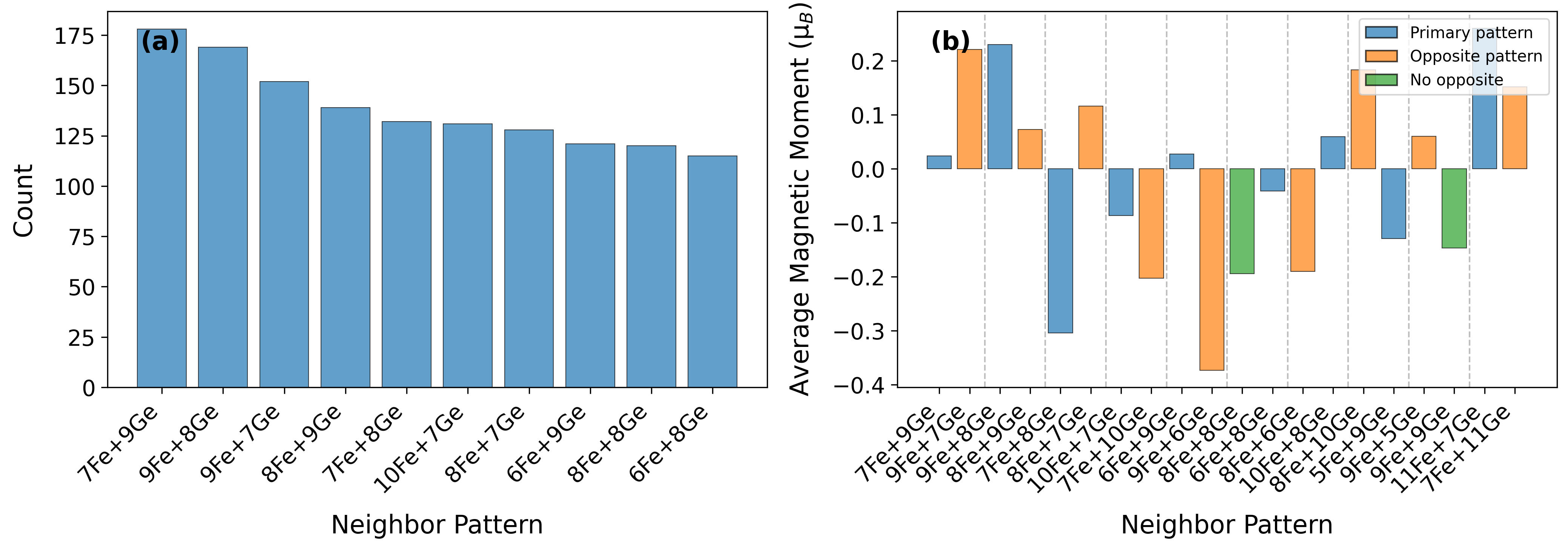}
\caption{Top ten common patterns surrounding Fe atoms (a) and correlation between the average magnetic moment of Fe atoms with its surrounding neighbor patterns (b)}
\end{figure*}

Now, let us look at the most common neighbor patterns within the 3.5\AA~radius spheres across all structures, as reported in Fig.~S5(a). The most common pattern was found to be 7 Fe and 9 Ge atoms, in total 16 atoms, followed by 9 Fe and 8 Ge, 9 Fe and 7 Ge, and so on, overall totaling 88 different patterns. Next, we looked at how different local environments or neighbor patterns affect the average magnetic moment of Fe atoms through magnetic and structural anisotropies and show it in Fig.~S5(b). However, we did not find a clear correlation (it is reported more thoroughly in the next section), but we found that the sign of the average magnetic moment of Fe flips between some of the opposite neighbor patterns but not all of them. For example, for patterns 7Fe+8Ge and 8Fe+7Ge, the sign of the average magnetic moment is opposite, but for patterns 7Fe+9Ge (most common) and 9Fe+8Ge (second most common), the signs are the same.

\subsection*{Statistical analysis on the magnitude and the sign of the local magnetic moments}

In order to get a more comprehensive understanding of how local motifs or environments surrounding Fe atoms affect their magnetic moments, we report their dependence on the number of Fe neighbors, the number of Ge neighbors, the number of total neighbors, average Fe--Fe distance, and average Fe--Ge distance in Fig.~S6(a-e). In addition, we show the correlation coefficients between the magnetic moment with Fe, Ge, and total neighbors in a 4x4 grid plot in Fig.~S6(f). Our results show no correlation between magnetic moment and other parameters; however, one can see a mirror symmetry at 0 $\mu_B$, in each plot in panels (a-e), indicating that the parameters have no effect on the sign of magnetic moments of Fe atoms. Indeed, if we take the absolute value of the magnetic moment and plot its distribution as a function of the number of Fe and Ge neighbors, shown in Fig.~S7, we find (almost) a linear dependence between them. Our results show that the magnitude of the magnetic moments of Fe atoms increases (almost linearly) with the number of Fe neighbors but decreases with a greater number of Ge atoms. We fitted linear regression to our results, and correlation coefficients, \textit{r}, are 0.432 and -0.216 for the number of Fe and Ge neighbors, showing a reverse effect from them on the magnitude of the magnetic moment.

\begin{figure*}
\includegraphics[width=0.8\textwidth]{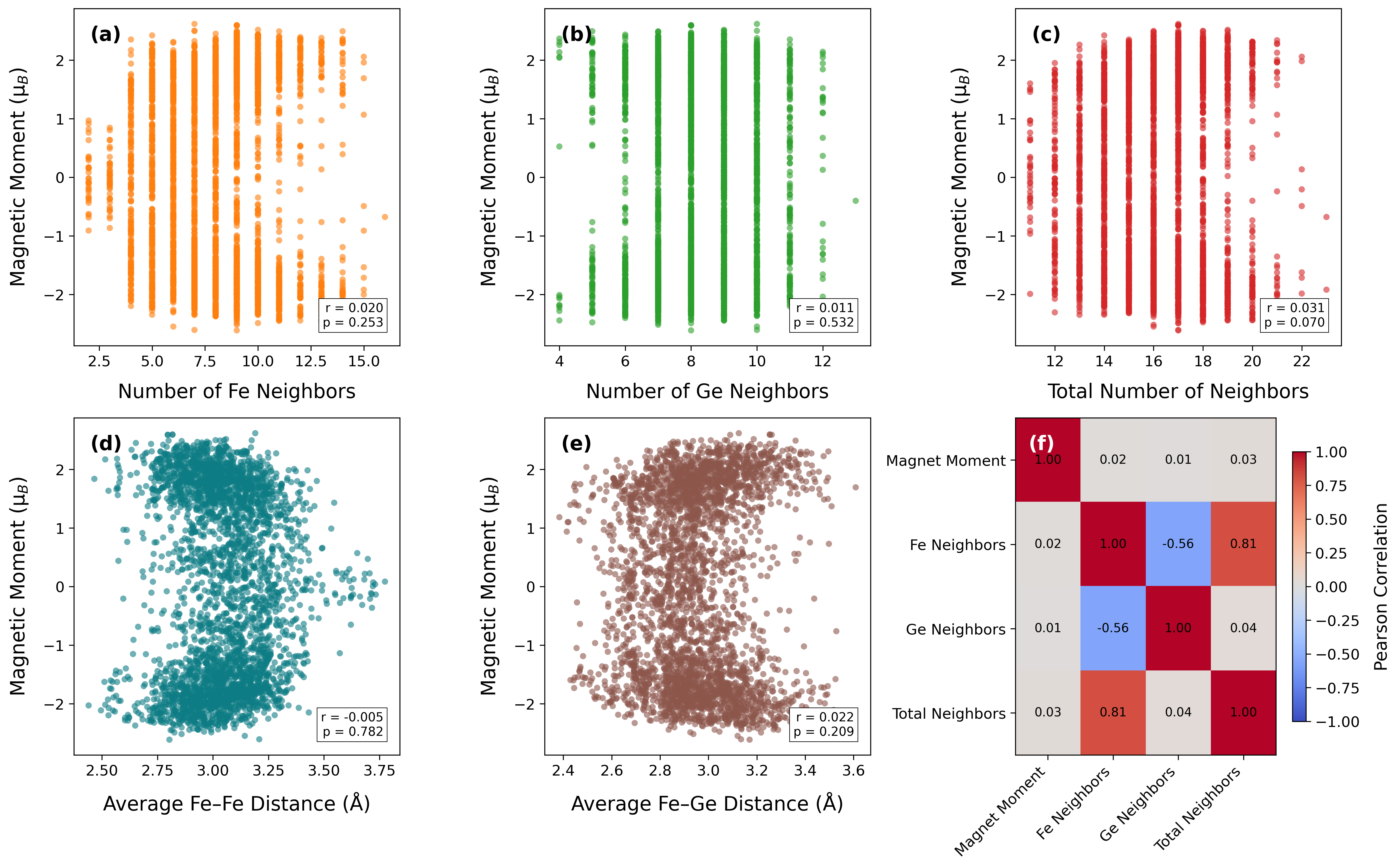}
\caption{Magnetic moments of Fe atoms as a function of the number of Fe neighbors (a), the number of Ge neighbors (b), the number of total neighbors (c), average Fe--Fe distance (d), and average Fe--Ge distance(e) are shown here with a correlation matrix between them shown in panel (f).}
\end{figure*}

\begin{figure*}
\includegraphics[width=0.8\textwidth]{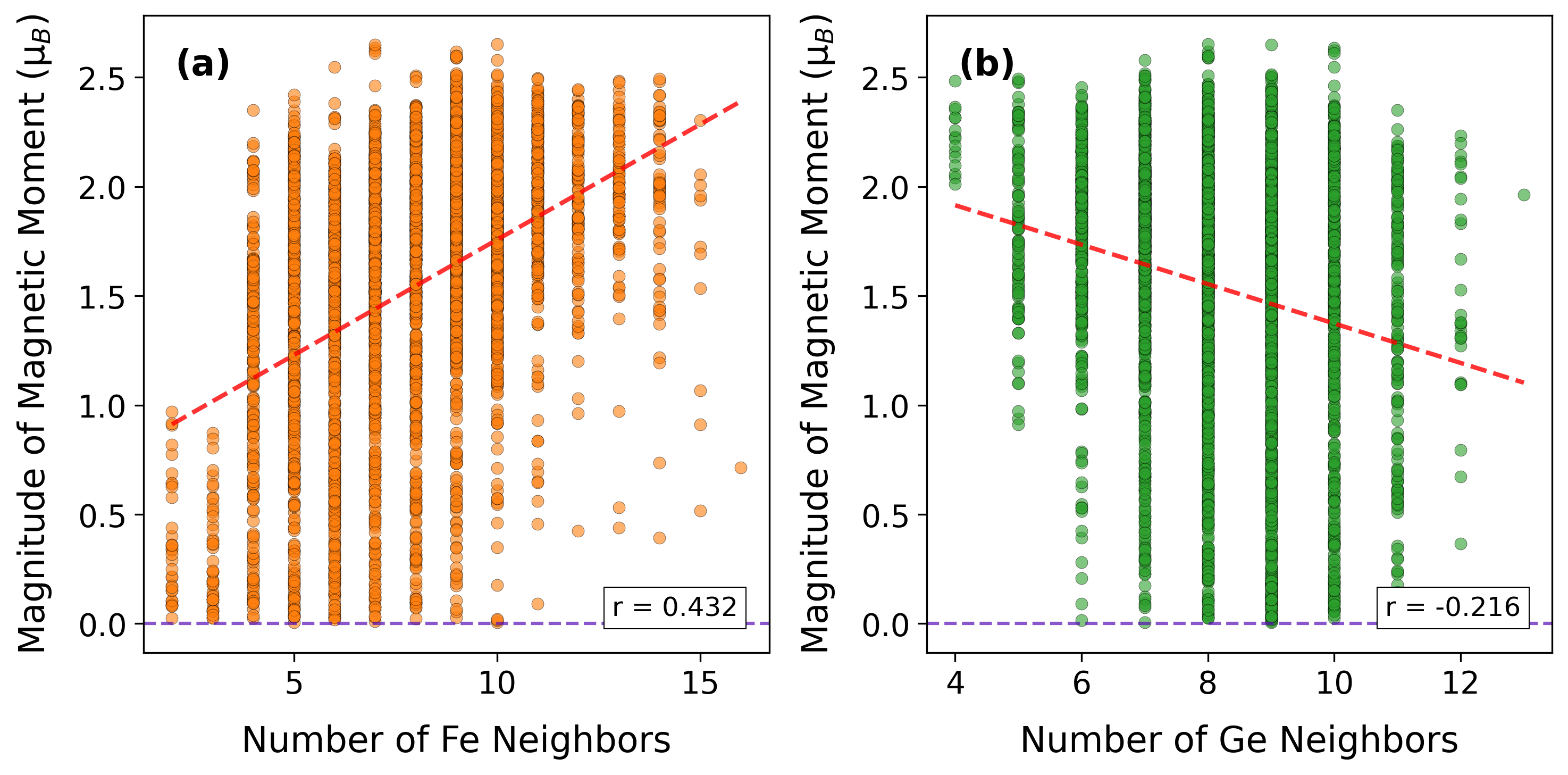}
\caption{The magnitude of magnetic moments of Fe atoms as a function of the number of Fe neighbors (a) and the number of Ge neighbors (b).}
\end{figure*}

Now, let us find out the reason for magnetic moments having both positive and negative signs. From Fig.~S5(b), there is an indication that the sign of the magnetic moment has some correlation with the local motifs. Thus, the sign of the magnetic moment could be dependent on the following three properties: (i) bond angles; (ii) local structural motifs; and (iii) the surrounding magnetic environment or mean field around Fe atoms. 

For the bond angle study, we report the results in Fig.~S8, where we show the bond angle distribution by the sign of magnetic moment, magnetic moment as a function of bond angle, and 2D density plot of the magnetic moment as a function of bond angle for Fe--Fe--Fe, Fe--Fe--Ge, and Ge--Fe--Ge bonds, respectively. Reported magnetic moments are from the Fe atoms at the center of bonds. The distribution of bond angles shows that in an amorphous system, one can expect to have complex and varying bond angles ranging from 40\degree to 180\degree. Interestingly, amorphous FeGe structures were found to have the majority of their bond angles similar to the crystalline counterpart, with peaks of histograms overlapping with the reference lines from the crystalline B20 structure, especially for Fe--Fe--Fe bonds. However, as shown in Fig.~S8(d-f), the sign and magnitude of the magnetic moment of the Fe atoms were found to have no correlation with the bond angles.

\begin{figure*}
\includegraphics[width=0.8\textwidth]{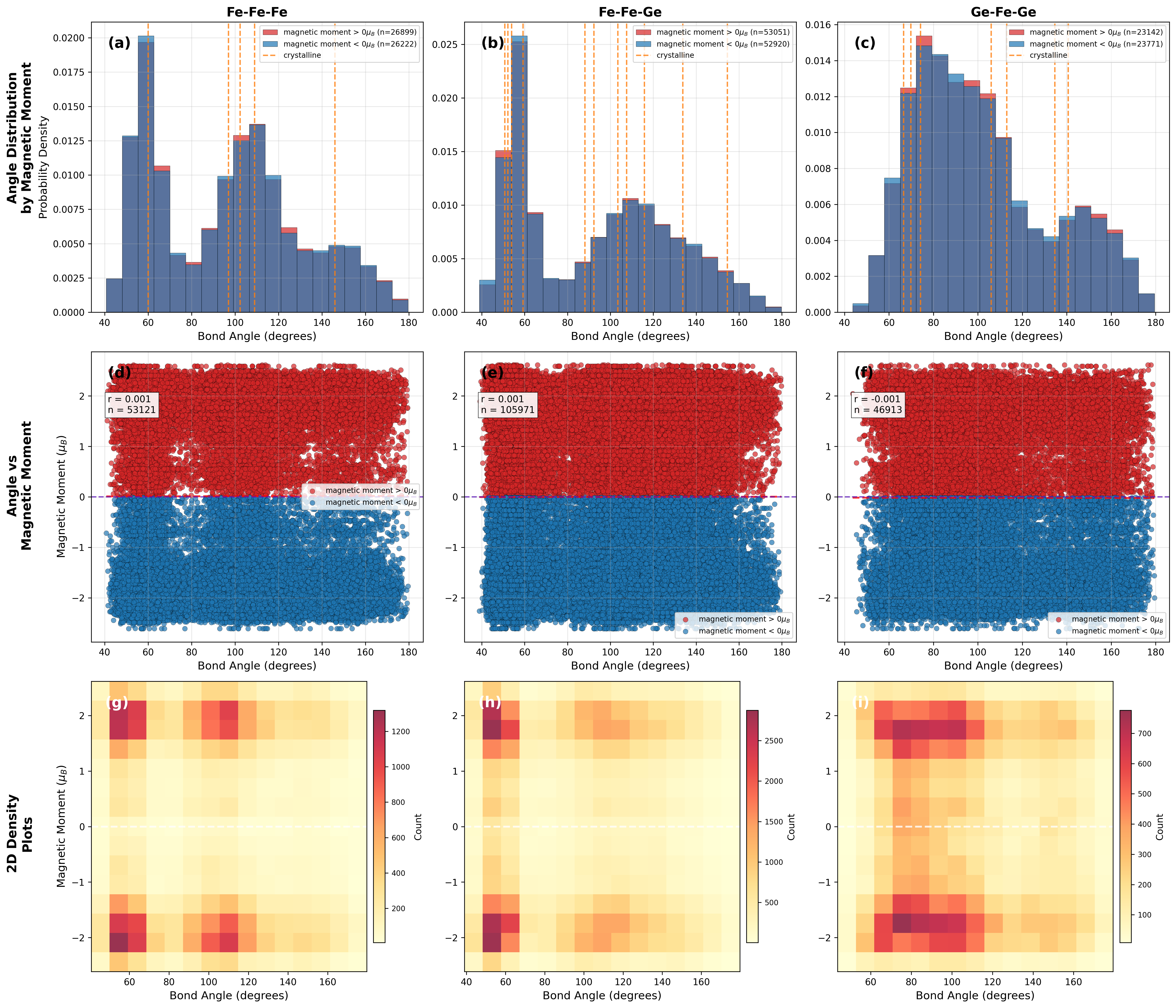}
\caption{Distribution of bond angles colored by the sign of magnetic moment (a-c), magnetic moment as a function of bond angle (d-f), 2D density plot of the magnetic moment as a function of bond angle (g-i) for Fe--Fe--Fe, Fe--Fe--Ge, and Ge--Fe--Ge bonds, respectively. Vertical dashed lines are the bond angles from the crystalline B20 structure.}
\end{figure*}

\begin{figure*}
\includegraphics[width=1\textwidth]{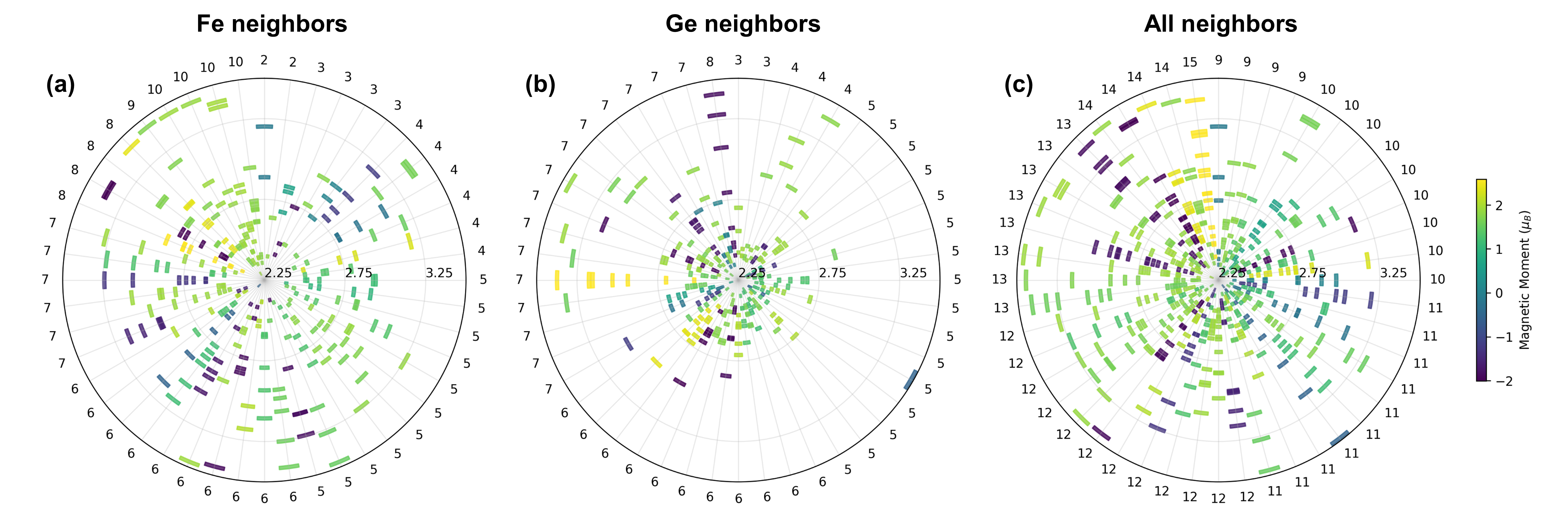}
\caption{Barcode representation of the local motifs in amorphous FeGe for Fe (a), Ge (b), all neighbors (c), respectively. Motifs are shown in an order where the number of neighbors increases in a clockwise direction. Center of the sphere starts from the radius of 2.25\AA~to 3.5\AA. Coloring is based on the magnetic moment of Fe atoms in the center.}
\end{figure*}

\begin{figure*}
\includegraphics[width=0.8\textwidth]{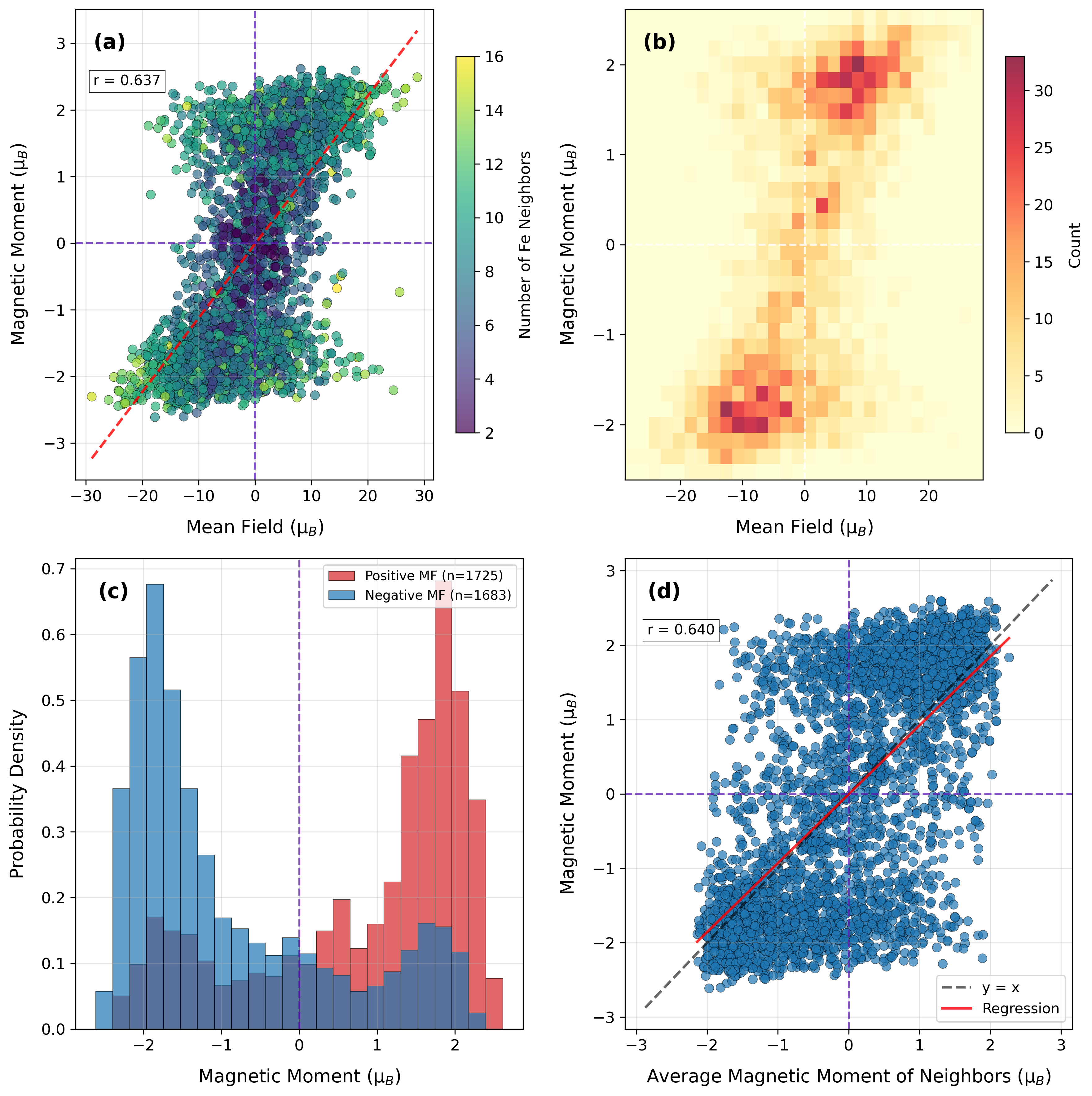}
\caption{The effect of local mean field on the magnetic moments of Fe atoms. Magnetic moments as a function of the surrounding mean field (a), density plot of mean field (b), distribution of the sign of the mean field as a function of magnetic moment (c), and magnetic moment as a function of the averaged magnetic moments of the neighbors. The mean field is taken as the sum of the magnetic moments of surrounding Fe neighbors within a 3.5\AA~sphere. Linear regression is fitted with the corresponding Pearson correlation coefficient.}
\end{figure*}

For the study of local structural motif influence on the magnetic moments, we developed a barcode method and report the results in Fig.~S9. The barcode method works as follows: we expand a sphere centered on Fe atoms with a radius of 2.25\AA~ to 3.5\AA~and draw a bar when the sphere reaches a neighbor. The bars are wider if the sphere touches multiple neighbors at the same time, and the bars are colored based on the magnetic moment of the center Fe atoms. For visual clarity, we chose the structure in which we have calculated the magnetic exchange coupling coefficients and show the barcode representation of the local motifs of that structure in Fig.~S9. Our results show that there is no direct correlation between local structural motifs and the sign of the magnetic moment of Fe atoms. Note that the results are qualitatively the same across all the amorphous structures in Fig.~1(b).

Lastly, we look at the effect of the surrounding magnetic environment, or the mean field, on the magnetic moment of the Fe atoms and report the results in Fig.~S10. Fig.~S10(a) shows the magnetic moments of the Fe atoms and the mean field values surrounding them. Panel(b) shows the 2D density of the results in panel (a). Here, our results show that the sign and the magnitude of the magnetic moments depend on the surrounding mean field and have an almost linear correlation between them. Similarly to what we reported in Fig.~S7, the magnitude of the magnetic moment is found to be dependent on the magnitude of the mean field. Additionally, the sign of the magnetic moment is also found to be dependent on the sign of the mean field surrounding the magnetic ions in the amorphous magnets, in this case, Fe atoms. We fit a linear correlation to our data and found the correlation function, r, to be 0.637. In addition, we report the distribution of the sign of the mean field as a function of the magnetic moment in Fig.~S10(c). Our results show that the majority of the magnetic moments tend to have the same sign as the mean field surrounding them; however, we also find some of them having the opposite sign to the mean field. These results show that amorphous magnets have complex magnetic interactions, and there can be regions with antiferromagnetic interactions within ferromagnetic amorphous systems which was found to be true when we calculated the magnetic exchange interaction coefficients reported in Fig.~3 and Fig.~S11. Next, we show the magnetic moments of Fe atoms as function of the average magnetic moment of surrounding neighbors by dividing the mean field by the number of neighbors in Fig.~S10(d). These results also show the linear dependence of the magnitude and the sign of the magnetic moments of Fe atoms. In this case, after a linear correlation fit, r is 0.640.   

\section*{Supplementary Note 3: Bond-resolved magnetic interactions in amorphous FeGe}

\subsection*{Exchange interactions}

The calculated exchange couplings span a remarkably broad range in both magnitude and sign, from strongly ferromagnetic ($-555$~meV) to antiferromagnetic ($+84$~meV) (Figs.3(a) and S11(a)).  
This distribution is nearly two orders of magnitude wider than in crystalline B20 FeGe, where $|J_{ij}|\!\sim\!8$~meV~\cite{muhlbauer2009}.  
The distribution is asymmetric, with approximately 75\% of the couplings ferromagnetic and 25\% antiferromagnetic (Fig.~3(e)), consistent with the experimentally observed net ferromagnetism in amorphous FeGe.  
The mean and median exchange values are $-109$ and $-76$~meV, respectively.

A central result is that exchange interactions in amorphous FeGe cannot be described as a function of interatomic distance alone.  
Instead, we find that $J_{ij}$ depends jointly on the separation $r_{ij}$ and on the magnitudes of the interacting local moments $m_i$ and $m_j$,
\begin{equation}
J_{ij} = J(r_{ij}, m_i, m_j).
\end{equation}
This dependence arises because the local moment amplitudes fluctuate strongly with coordination environment, modifying the degree of $d$-electron overlap and exchange splitting that mediate magnetic coupling.  
Short Fe–Fe bonds connecting strongly magnetic sites tend to support robust ferromagnetic exchange, whereas longer or Ge-mediated pathways—often involving reduced local moments—yield weaker or antiferromagnetic interactions (Figs.3(a) and S11(a)).

To give more perspective, we show J values as 3D scatter plots that depend on the magnetic moments of the pairs and their distance from each other in Fig.~S12(a).

Earlier first-principles studies of amorphous alloys reported substantial scatter around a single $J(r)$ curve~\cite{terasawa2020,Lindmaa2013}, but did not explicitly treat moment amplitude as an independent control parameter.

Here, we fitted $J$ using multivariate polynomial regression $J_{fit} = \sum_{i=0}^{N} \sum_{j=0}^{N-i} c_{ij} r^i \mu^j$, where $r$ is the pair distance and $\mu$ is the product of the magnetic moments of the pairs, and found that the model fits well to the $J$ values with $R^2$ reaching 0.68 with 4th-order polynomial degree. This confirms that moment heterogeneity provides an essential additional axis governing exchange variability. Higher-order polynomials were found to overfit even with higher $R^2$ numbers, and this polynomial model was considered following the work of Lindmaa \textit{et al.}\cite{Lindmaa2013} as our $J$ distribution as a function of r resembles their $J$ distribution using the magnetic Connolly-Williams\cite{CW1983} approach.
We emphasize that this fit serves as a descriptive parameterization rather than a microscopic model of exchange.

The joint dependence of $J_{ij}$ on geometry and moment magnitude represents a fundamental departure from conventional models of metallic glasses, which assume a distance-only Bethe–Slater form $J=J(r)$~\cite{coey1978}.  
In amorphous FeGe, local-moment fluctuations are comparable in scale to structural disorder, rendering the distance-only Heisenberg picture insufficient.

To elucidate the origin of the exchange-sign distribution, we analyzed the local magnetic environment of each Fe atom.  
The sign of the local moment correlates nearly linearly with the mean magnetic field generated by neighboring spins (Fig.~2(c)), indicating a strong local mean-field alignment tendency.  
This explains the predominance of ferromagnetic exchange and the observed net magnetization, while isolated antiparallel moments occur at interfaces between locally ferromagnetic regions, acting as precursors to domain-wall-like structures, unlike spin glasses, which are formed in other amorphous systems.

Regions with increased Fe concentration in amorphous Fe$_x$Ge$_{1-x}$ result in enhanced local Fe coordination and moment amplitude (Fig.~2(a)), thereby strengthening exchange interactions; consistent with this trend, a ferromagnetic ground state is observed experimentally for $x>0.63$~\cite{streubel2021}.

\begin{figure*}
\includegraphics[width=1\textwidth]{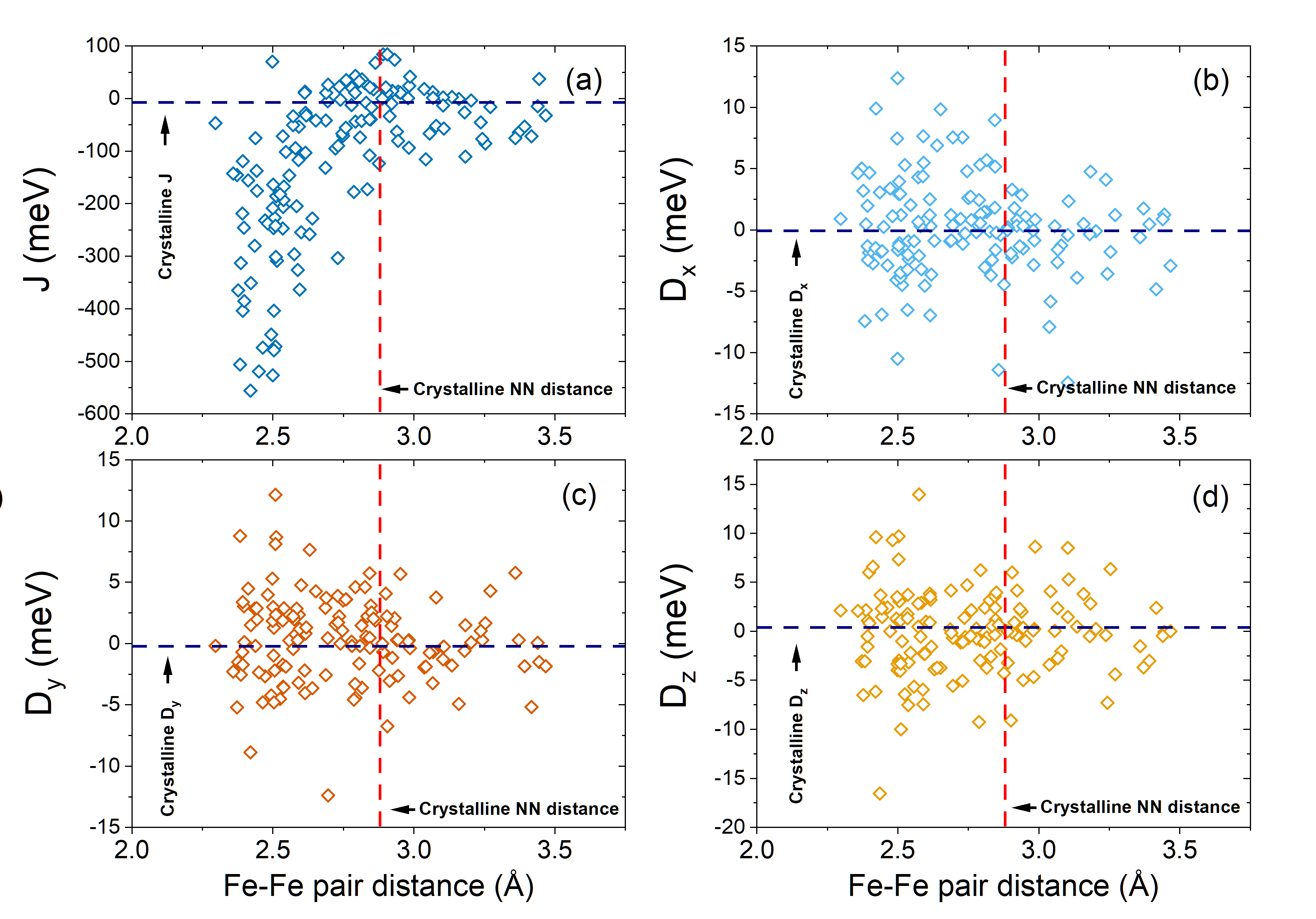}
\caption{Calculated magnetic exchange couplings (a) and three components of the DMI (b-d) of the amorphous FeGe structure are shown as a function of the Fe--Fe pair distances. The horizontal line references the corresponding 1NN values of the crystalline FeGe structure in Fig.~S1, and the vertical line represents the 1NN bond-length of the crystalline FeGe.}
\end{figure*}

\begin{figure*}
\includegraphics[width=1\textwidth]{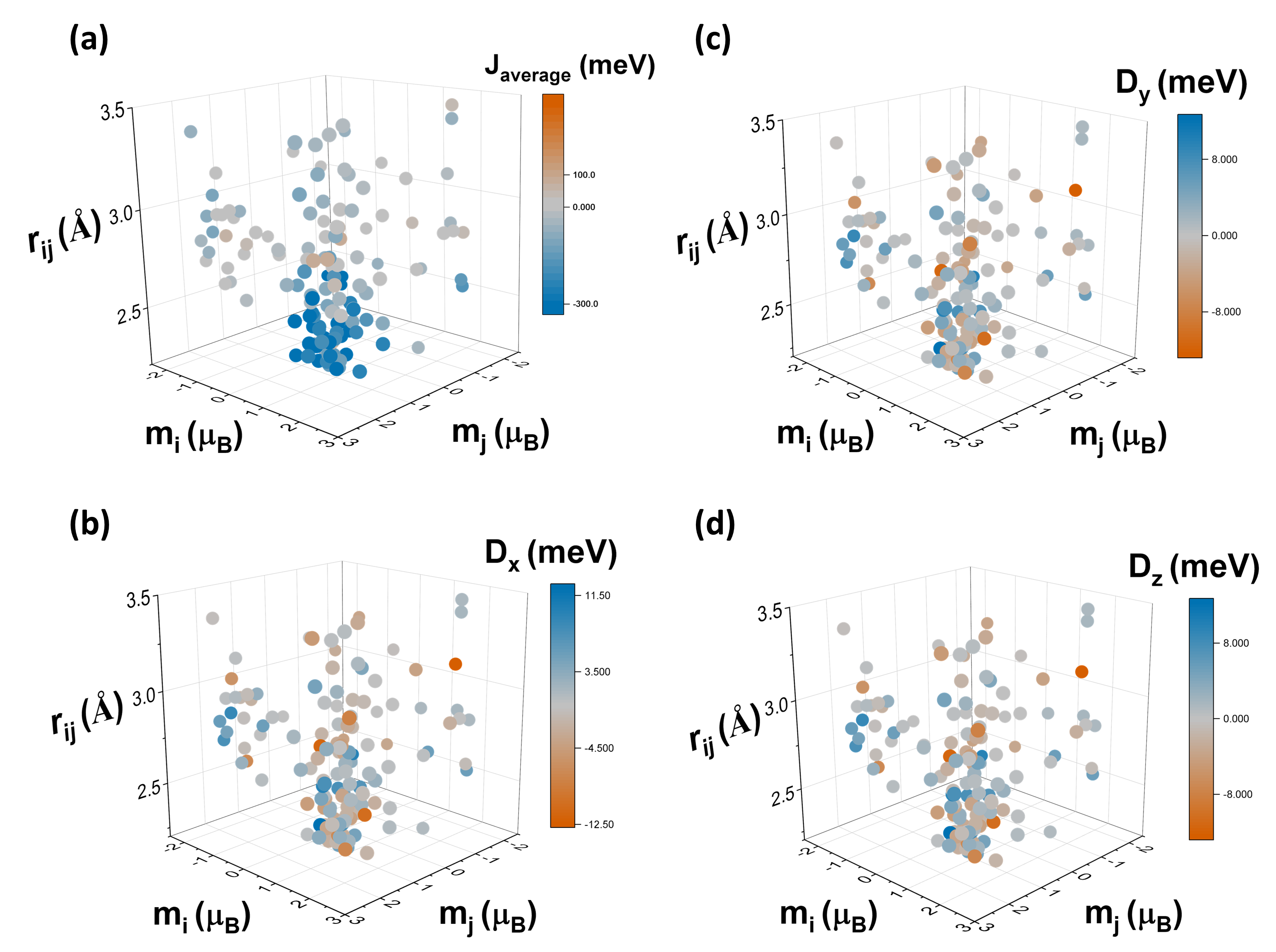}
\caption{Calculated magnetic exchange couplings (a) and three components of the DMI (b-d) of the amorphous FeGe structure are shown as a 3D scatter plot for a better viewing angle and represent the idea more clearly. The x and y axes are the magnetic moments of the Fe--Fe pairs, with their distances being shown along the z-axis.}
\end{figure*}

\subsection*{Dzyaloshinskii-Moriya interactions}

The DMI vectors exhibit a similarly broad distribution (Fig.~3(b–d) and S11(b-d)), with magnitudes reaching up to 16.6~meV—significantly larger than the crystalline value of $\sim$0.5~meV.  
While the mean (0.163, 0.370, -0.016~meV) and median (0.250, 0.274, 0.004 ~meV) DMI vectors magnitudes remain comparable to those in crystalline FeGe, the wide spread reflects strong local symmetry breaking and chemical disorder in the amorphous phase.
As with exchange, the DMI depends on both pair distance and local moment amplitude, contradicting the assumption of a purely distance-dependent interaction. Fig.~S12(b-d) is to provide more perspective that DMI is also a function of pair distance and moments of the pairs, same as J.

\subsection*{Single Ion Anisotropy}

The calculated single-ion anisotropy eigenvalues also span a broad range, with maxima up to 1.76~meV (Fig.~3(f)).  
Interestingly, in some SIA matrices, the off-diagonal elements were found to be larger than the diagonal elements; thus, we diagonalized the SIA matrices, and we report the maximum absolute eigenvalues from the diagonalization in Fig.~3(f).
Diagonalization of the local anisotropy tensors reveals that the largest eigenvalues increase systematically with local coordination, indicating that denser environments promote stronger anisotropy.  
This trend is consistent with experimental phase diagrams of amorphous FeGe alloys, where increasing Fe concentration drives transitions from disordered to helical and ferromagnetic phases~\cite{streubel2021}.

Furthermore, we studied the effect of the local crystal environment on the anisotropies each magnetic ion is subject to. To do so, we utilized the barcode method (see Supplementary Note 5) mentioned above and reported the results in Fig.~S14. Fig.~S14 shows the barcode representations, and they are colored by the maximum absolute eigenvalue of SIA of each Fe atom in the studied amorphous structure. Fig.~S14(c) shows the barcode representation of the local environment from all the neighbors for each Fe atom, and one can see that, in a qualitative way, coloring changes from blue to red when the number of neighbors increases or goes in a clockwise direction. In other words, our results show that Fe atoms are subject to more anisotropy with increasing number of total neighbors, not neighbor counts by the individual atom types, see Fig.~S14(a-b).

\subsection*{Super-exchange}

\begin{figure*}
\includegraphics[width=0.8\textwidth]{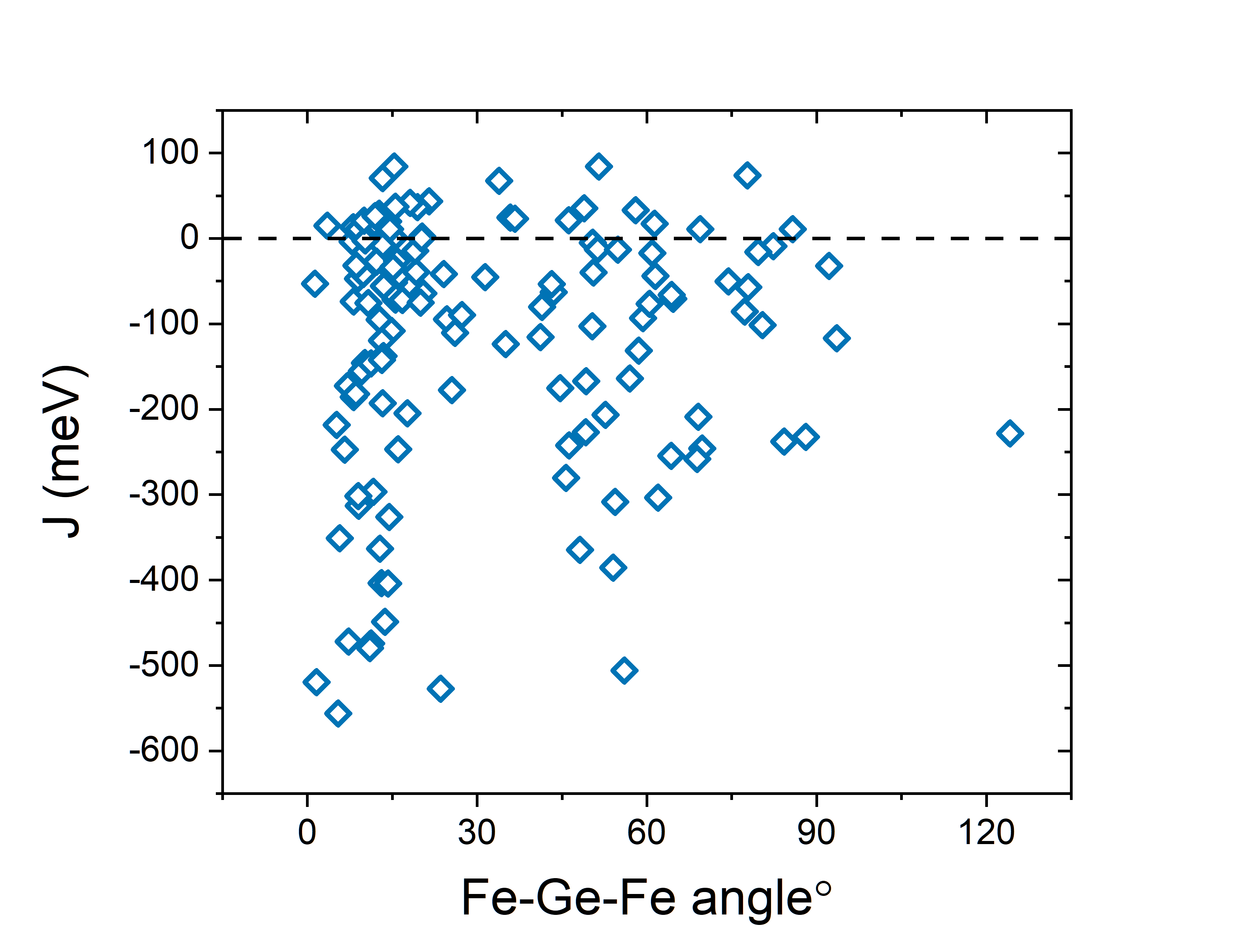}
\caption{Exchange coupling interactions, J, as a function of the Fe--Ge--Fe angle.}
\end{figure*}

Next, to check whether there is a superexchange interaction in amorphous FeGe, we take into account the Fe--Ge--Fe angle by finding the closest Ge ion between all the Fe pairs we studied to get their interaction parameters, and the results are shown in Fig.~S13. J as a function of the Fe--Ge--Fe angle is plotted. One can see that there is no linear correlation between J and the Fe--Ge--Fe angle. In fact, metallic amorphous systems should have either direct exchange coupling or indirect RKKY interactions, as mentioned in Ref. \cite{coey1978}.

\begin{figure*}
\includegraphics[width=1\textwidth]{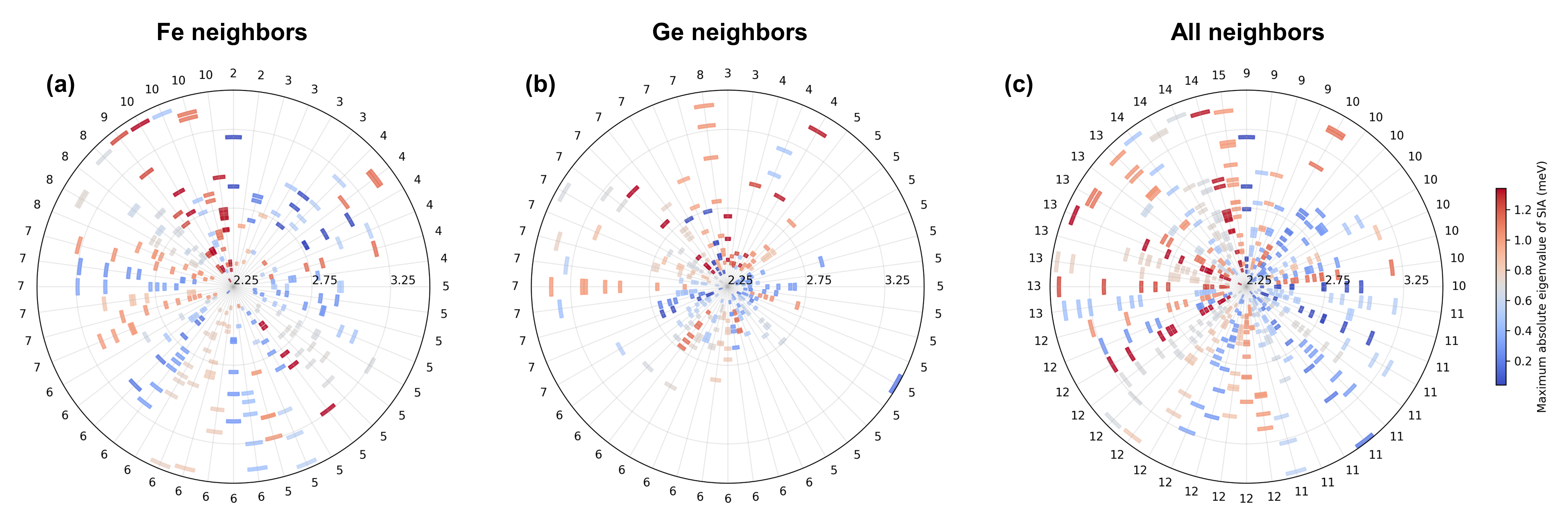}
\caption{Barcode representation of the local motifs in amorphous FeGe for Fe (a), Ge (b), all neighbors (c), respectively. Motifs are shown in an order where the number of neighbors increases in a clockwise direction. Center of the sphere starts from the radius of 2.25\AA~to 3.5\AA. Coloring is based on the maximum absolute eigenvalue of the SIA of Fe atoms in the center.}
\end{figure*}

\section*{Supplementary Note 4: Construction of spin Hamiltonian}
\textcolor{black}{
To characterize the magnetic exchange interactions in amorphous magnets, we take the most general $3\times3$ matrix form of the $J$ matrices as
\[ J = 
\begin{pmatrix}
J_{xx} & J_{xy} & J_{xz} \\
J_{yx} & J_{yy} & J_{yz} \\
J_{zx} & J_{zy} & J_{zz} 
\end{pmatrix}. \]
Such matrices can be decomposed into symmetric and anti-symmetric parts, which mainly correspond to the collinear spin alignments and DMI, respectively. The symmetric $J_{sym}$ is given by
\[J_{sym} =
\begin{pmatrix}
J_{xx} & \frac{1}{2}(J_{xy} + J_{yx}) & \frac{1}{2}(J_{xz} + J_{zx}) \\
\frac{1}{2}(J_{yx} + J_{xy}) & J_{yy} & \frac{1}{2}(J_{yz} + J_{zy}) \\
\frac{1}{2}(J_{zx} + J_{xz}) & \frac{1}{2}(J_{zy} + J_{yz}) & J_{zz} 
\end{pmatrix}.
\]
In crystalline or highly symmetric systems, $J_{sym}$ matrices prefer spins to be collinearly aligned and have negligible off-diagonal elements. In the amorphous case, we numerically found that the off-diagonal elements of $J_{sym}$ are also negligible compared to the diagonal part; thus, we only focus on \( J_{\alpha \alpha} \) (\( \alpha = x, y, \) and \( z \)) and report their average values in the manuscript. Note that we report the average value of the diagonal elements since all three elements are found to slightly vary from each other, as shown in the example below. 
\[J_{sym} (meV) =
\begin{pmatrix}
-234.342 &	-0.102 &	-0.070 \\
-0.102 &	-235.054 &	-0.261 \\
-0.070 &	-0.261 &	-235.331
\end{pmatrix}.\]
The antisymmetric part, $J_{asym}$, of the $J$ matrices is related to the DM interactions. Its relation to the DMI vector \textbf{D} can be obtained as follows:
\[
J_{asym} = 
\begin{pmatrix}
0 & \frac{1}{2}(J_{xy} - J_{yx}) & \frac{1}{2}(J_{xz} - J_{zx}) \\
\frac{1}{2}(J_{yx} - J_{xy}) & 0 & \frac{1}{2}(J_{yz} - J_{zy}) \\
\frac{1}{2}(J_{zx} - J_{xz}) & \frac{1}{2}(J_{zy} - J_{yz}) & 0 
\end{pmatrix} = \textbf{D} =  
\begin{pmatrix}
    0 & D_z & D_y \\
    D_z & 0 & D_x \\
    D_y & D_x & 0
\end{pmatrix}.
\]
$J_{asym}$ or $\textbf{D}$ favors the spins to be perpendicular to each other within the plane that has its normal vector parallel to $\textbf{D}$.
Furthermore, to study the single-ion anisotropy, we take into account the general $3\times3$ matrix form for the $A$ matrices:
\[ A = 
\begin{pmatrix}
A_{xx} & A_{xy} & A_{xz} \\
A_{yx} & A_{yy} & A_{yz} \\
A_{zx} & A_{zy} & A_{zz} 
\end{pmatrix}.
\]
All elements in $J_{sym}$, $J_{asym}$, and $A$ matrices are extracted using the four-state energy mapping method \cite{Xiang2013,Xu2019,Bayaraa2021a}. }

\section*{Supplementary Note 5: Barcode Method}

We implemented the barcode method to study the local environment around the Fe ions in the amorphous structures. We expand a sphere with a radius ranging from 2.25\AA~ to 3.5\AA~, following the analysis of the first peak of g$_{Fe--Fe}$ and g$_{Fe--Ge}$ in Fig.~S3(b-d), and indicate a neighbor by drawing a bar when the sphere reaches them. Bars are broader if the sphere reaches multiple neighbors at the same time. Compared to the standard structural fingerprint methods (i.e., local coordination environment, Voronoi decomposition, and classical force-field-inspired descriptors), this method provides more information on how neighboring atoms are distributed in a complex local environment, such as amorphous systems.

\newpage
\bibliography{library}